\documentclass[a4paper,twocolumn,11pt,accepted=2023-09-28]{quantumarticle}
\pdfoutput=1
\usepackage[utf8]{inputenc}
\usepackage[english]{babel}
\usepackage[T1]{fontenc}
\usepackage{amsmath}
\usepackage{amssymb}
\usepackage{nicefrac}
\usepackage{complexity}
\usepackage[numbers]{natbib}
\usepackage{xurl}
\usepackage{hyperref}

\usepackage{tikz}
\usetikzlibrary{quantikz}

\usepackage{caption}
\usepackage{subcaption}

\usepackage{ulem} 

\newtheorem{theorem}{Theorem}

\begin{document}

\title{Quantum circuit compilation and hybrid computation using Pauli-based computation}

\author{Filipa C. R. Peres}
\affiliation{International Iberian Nanotechnology Laboratory (INL), Av. Mestre Jos\'{e} Veiga, 4715-330 Braga, Portugal}
\affiliation{Departamento de F\'{i}sica e Astronomia, Faculdade de Ci\^{e}ncias, Universidade do Porto, rua do Campo Alegre s/n, 4169–007 Porto, Portugal}
\email{filipa.peres@inl.int}
\orcid{0000-0003-2465-4546}

\author{Ernesto F. Galvão}
\affiliation{International Iberian Nanotechnology Laboratory (INL), Av. Mestre Jos\'{e} Veiga, 4715-330 Braga, Portugal}
\affiliation{Instituto de F\'{i}sica, Universidade Federal Fluminense, Avenida General Milton Tavares de Souza s/n, Niter\'{o}i, Rio de Janeiro 24210-340, Brazil}
\orcid{0000-0003-2582-9918}

\twocolumn[
  \begin{@twocolumnfalse}
    \maketitle
    \begin{abstract}
    Pauli-based computation (PBC) is driven by a sequence of adaptively chosen, non-destructive measurements of Pauli observables. Any quantum circuit written in terms of the Clifford+$T$ gate set and having $t$ $T$ gates can be compiled into a PBC on $t$ qubits. Here we propose practical ways of implementing PBC as adaptive quantum circuits and provide code to do the required classical side-processing. Our schemes reduce the number of quantum gates to $O(t^2)$ (from a previous $O(t^3 / \log t)$ scaling) and space/time trade-offs are discussed which lead to a reduction of the depth from $O(t \log t)$ to $O(t)$ within our schemes, at the cost of $t$ additional auxiliary qubits. We compile examples of random and hidden-shift quantum circuits into adaptive PBC circuits. We also simulate hybrid quantum computation, where a classical computer effectively extends the working memory of a small quantum computer by $k$ virtual qubits, at a cost exponential in $k$. Our results demonstrate the practical advantage of PBC techniques for circuit compilation and hybrid computation.
    \end{abstract}
  \end{@twocolumnfalse}
  ]

\section{Introduction} \label{sec: Introduction}
Quantum computers promise to solve some computational tasks much faster than their classical counterparts. Such tasks include integer factoring~\cite{Shor94}, dynamical quantum simulation~\cite{Lloyd96}, solving systems of linear equations~\cite{HHL09}, and certain problems in graph, group and knot theory~\cite{Montanaro16}. These algorithms demonstrate how large-scale, error-corrected quantum computers are expected to unlock several practical applications. However, the currently available devices--dubbed NISQ, an acronym coined by John Preskill standing for noisy intermediate-scale quantum~\cite{Preskill18}--have several limitations such as a limited number of qubits and qubit connectivity, and a considerable amount of coherent and incoherent errors.

While some impressive results have recently been demonstrated, with quantum computational advantage being achieved for non-trivial (albeit contrived) mathematical tasks~\cite{Arute19, Zhong20, Wu21-supremacy}, the limitations of available devices preclude any serious practical implementation of the promising aforementioned algorithms. Therefore, and since fault-tolerant quantum computation might still be decades away, it makes sense to ask ourselves how we can best exploit the currently available NISQ technology to reach computational advantage for practical applications in the near term.

Due to the constraints of such devices, it is clear that any reasonable implementation must make use of small qubit counts and shallow circuits. Such requirements can be achieved by developing smart schemes wherein a classical computer and a small quantum processing unit (QPU) are combined to make the best use of the limited quantum resources. Some key results are hybrid quantum-classical algorithms for ground-state energy estimation of a quantum Hamiltonian using the variational quantum eigensolver algorithm~\cite{Peruzzo14}, speeding up the solution to 3-satisfiability problems~\cite{Dunjko18}, and Bayesian inference using large data sets~\cite{Harrow20}. 

In line with this idea of reducing quantum resources, here we explore Pauli-based computation (PBC)~\cite{BSS16}, a model that uses a sequence of adaptively chosen, non-destructive measurements of Pauli observables to do universal quantum computation. We explore PBC as a compilation tool that can be used to trade (affordable) classical computation for quantum computation, but also as a natural framework to formulate a hybrid computation set-up wherein a classical (super)computer and a small quantum computer can be used to emulate a larger quantum machine.

More specifically, here we focus on the task of using PBC to compile large $n$-qubit Clifford$+T$ quantum circuits, with $t$ $T$ gates, into (often) less resource-demanding adaptive circuits that can be run on quantum hardware, using control and side-processing of a classical computer~\cite{BSS16}. We start by presenting three different theoretical schemes for translating the sequence of non-destructive measurements of pairwise commuting $t$-qubit Pauli operators of a standard PBC back into the quantum circuit model in an efficient and practical way. Using the first of these schemes, the PBC can be carried out with $O(t^2)$ gates and depth, while the second and third schemes require a gate count of the same order but have a reduced circuit depth of $O(t \log t)$ and $O(t)$, respectively. These results improve upon the proposal found in Ref.~\cite{Yoga19} which is not only slightly more involved but also requires a number of gates that scales like $O(t^3 / \log t)\,.$ Besides these theoretical contributions, we also developed Python code to implement the efficient PBC circuit compilation we have just described. This code is available at \url{https://github.com/fcrperes/CompHybPBC} and can be used to perform two distinct tasks. In both tasks, we use \textit{PBC compilation} to provide us with equivalent PBC circuits which often have lower depth and/or gate counts, and, under certain conditions, a smaller number of qubits than the corresponding circuit input to the code. The first task that can be performed is that of weak simulation of the original circuit. The second task consists of carrying out approximate strong simulation while simultaneously reducing the number of qubits required, by simulating some of them classically; we call these \textit{virtual qubits} and this particular task \textit{hybrid PBC}, following the terminology in Ref.~\cite{BSS16}.\footnote{Here, it is important to note that the first task--that of \textit{PBC compilation} together with the weak simulation of the original quantum circuit--is also a hybrid quantum-classical procedure in the sense that part of the computation is being carried out by the controlling classical computer. Even so, and in order to avoid possible confusion, the reader should keep in mind that here the term \textit{hybrid PBC} is exclusively used to denote the second task explored in this paper, i.e. that of reducing the number of qubits needed by the quantum hardware, by simulating a certain number of \textit{virtual qubits}.} This second task can be phrased in a slightly different (albeit equivalent) manner as a way of using a classical computer to extend the number of qubits available to the QPU by $k$ virtual qubits, with a cost exponential in $k$.

We expect this work to be relevant both in the near term and in the long term. Firstly, PBC compilation allows the efficient reduction of intermediate-sized quantum circuits into smaller instances that can be run on current or near-term quantum hardware. Secondly, this reduction of quantum resources (i.e. number of qubits, depth and/or gate counts) will remain relevant even when large-scale, fault-tolerant quantum computers become available. Finally, hybrid PBC might be of particular interest to supercomputing centers wherein the powerful, available classical machines can be used to extend the working memory of a small QPU.

This paper is organized as follows. We start by presenting a review of previous work (Section~\ref{sec: Preliminaries}). Then, in Section~\ref{sec: Main contributions}, we discuss the theoretical contributions of our research. In Section~\ref{sec: Numerical results}, we report a practical implementation of PBC via Python code and the corresponding numerical results achieved with it for a particular class of hidden-shift circuits and also for random quantum circuits. Finally, in Section~\ref{sec: Concluding remarks}, we draw some conclusions and set out possible future lines of research.

\section{Review of prior work} \label{sec: Preliminaries}
\subsection{Classical simulation of quantum circuits} \label{subsec: Classical simulation of quantum circuits}
There are two distinct notions of classical simulation. Weak simulation refers to the ability to sample from the (exact) same output distribution as that of the quantum circuit we intend to simulate. On the other hand, strong simulation consists of (exactly) computing the probability of obtaining a specific computational outcome. Additionally, approximate versions of each of these tasks can also be considered. In approximate weak simulation, we sample from an output distribution that is $\epsilon$-close to the output distribution of the original circuit with respect to the $\ell_1$-norm. In approximate strong simulation, we estimate the probability of a certain outcome within a given error $\epsilon$ that is either additive, multiplicative or relative, depending on the algorithm.\footnote{In the literature, the notion of \textit{multiplicative} errors is often used in a broad sense that encompasses also the so-called \textit{relative} errors.} For more details on the different types of errors see Ref.~\cite{PhDCalpin}.

The Gottesman-Knill theorem~\cite{PhDGottesman, Gott98} states that polynomial-sized, unitary Clifford circuits, i.e. circuits with only (i) stabilizer-state inputs, (ii) gates drawn from the Clifford set, and (iii) final measurements in the computational basis, are classically simulable in polynomial time (both in the weak and in the strong sense). On the other hand, by adding any non-Clifford single-qubit unitary to the available set of gates, universal quantum computation is unlocked. As stated in the introduction, quantum computers are strongly believed to be more powerful than classical computers. This means that, unlike unitary Clifford circuits, universal quantum circuits cannot be efficiently simulated by classical means.

Regardless, the ability to simulate increasingly larger universal quantum circuits is desirable because, on the one hand, it will allow us to validate initial fault-tolerant prototypes, and, on the other, it helps to establish the boundary between the tasks that can be classically computable and those that strictly require a quantum computer. Thus, there is a strong incentive to study ways of improving the exponential scaling of such classical simulators, making this a flourishing field of research.

Initial simulators had a runtime that scaled exponentially in the total number of qubits, $n$, of the quantum circuit. Since then, several different methods have been developed which have helped to improve this exponential factor. For example, using techniques based on tensor networks has given rise to simulators whose runtime scales exponentially not with the total number of qubits but rather with the treewidth of the underlying graph of the circuit~\cite{MarkovShi08}; more recently, but still within this framework, Huang,  Newman, and Szegedy have established a lower bound on the time complexity of strong simulation which goes as $2^{n-o(n)}$~\cite{Huang18}.

Another approach is to use quasi-probability methods. In this context, Pashayan, Wallman and Bartlett~\cite{Pashayan15} proposed a classical strong simulation method for qudits whose runtime depends on a measure of the total amount of negativity in the quantum circuit. More recently, in Ref.~\cite{RaussBV20}, a phase-space algorithm for classical simulation was proposed which extends to quantum systems of any (finite) dimension. This algorithm allows one to estimate the output probability of a computation performed by Clifford gates and Pauli measurements on negatively represented input states with a cost that is proportional to a robustness measure squared. This quantity has been proven to be smaller than or equal to the robustness of magic.

In the context of $n$-qubit Clifford circuits augmented by $t$ non-Clifford gates, Aaronson and Gottesman~\cite{AarGott04} showed that such circuits can be classically simulated in a time that scales like  $O(2^{4t}\poly(n))$. Afterward, stabilizer-rank approaches~\cite{BSS16, BG16, BBCCGH19, Qassim2019} have allowed improving the exponential scaling even further to $2^{\alpha t}$, where $\alpha < 1$, both in the context of strong and weak simulation. At the moment of writing, the best known coefficients are $\alpha = 0.228$ for  weak simulation~\cite{BG16, BBCCGH19} and $\alpha = 0.3963$ for strong simulation~\cite{Qassim21}.

Recently, Ref.~\cite{KissWet21} has combined the work done on stabilizer decompositions with tensor-network-based techniques to achieve enhanced strong classical simulation, allowing to successfully simulate random $50$- and $100$-qubit Pauli exponential circuits with up to $70$ $T$ gates, as well as $50$-qubit hidden-shift circuits with up to $1400$ $T$ gates.

In this paper, instead of considering the task of classically simulating a given universal quantum circuit, we approach the problem from the perspective of using a less resource-demanding quantum circuit to simulate a circuit with larger qubit counts, depth and/or gate counts. Within this framework, we focus on two different tasks.

We start by considering PBC compilation and the simulation of large circuits by circuits (often) requiring fewer quantum resources. Thus, the first obvious difference between this approach and the full classical simulation schemes described above is that, in our case, part of the computation is still executed on a quantum computer, while the other part consists of polynomial classical side-processing carried out by an assisting classical-control computer. As we will see, this procedure allows for the efficient (weak) simulation of the original circuit; the trade-off, however, is that the compiled quantum computation needs to be performed adaptively, i.e. interleaved with the classical processing. 

Secondly, oftentimes, the task of weak simulation of universal quantum circuits is translated into the problem of computing certain probabilities or marginals, which are then used to carry out the sampling. This is true of tensor-network and phase-space algorithms, as well as some stabilizer-rank-based simulators (e.g. Refs.~\cite{BSS16, BG16}). In contrast, our scheme executes a weak simulation of the original circuit without requiring the reduction to a strong simulation task. 

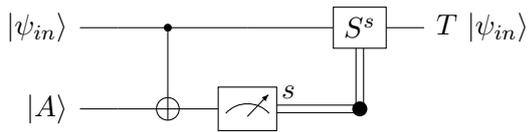
\begin{figure}
    \centering
    \begin{tikzpicture}
      \node[scale=1.0] {
        \tikzset{
            my label/.append style={above right,xshift=0.45cm,yshift=-0.25cm}
        }
        \begin{quantikz}[thin lines]
           \lstick{\ket{\psi_{in}}} & \qw & \ctrl{1} & \qw & \gate{S^s} & \qw \rstick{$T$ \ket{\psi_{in}}}\\
           \lstick{\ket{A}} & \qw & \targ{}  & \meter{$s$} & \cwbend{-1}
        \end{quantikz}      
      };
    \end{tikzpicture}
    \caption{Implementation of the $T$ gate via the well-known $T$-gadget.}
    \label{fig:T_gadget}
\end{figure}

Finally, compared against the weak simulators based on stabilizer-rank techniques proposed in Refs.~\cite{BG16} and \cite{BBCCGH19}, which allow to approximately sample from the output distribution of the desired circuit, our method provides an \textit{exact} weak simulation scheme.

The theoretical background behind this first scenario (which we call \textit{Task 1}) is discussed in Sections~\ref{subsec: Extended GK} and \ref{subsubsec: PBC - part 1}.

Additionally, we also consider a second task (dubbed \textit{Task 2}) wherein we reduce the demand on the number of qubits of the quantum hardware, by simulating some of the qubits classically (virtual qubits). The two main differences with respect to Task 1 are that (i) we now focus on the task of estimating the probability, $p$, that a given qubit yields the outcome 1, and (ii) a single estimate of $p$ involves running $2^{O(k)}$ different quantum circuits, where $k$ is the chosen number of virtual qubits.  We see that, in this case, and unlike Task 1, we need to run exponentially many circuits, so that this task is clearly \textit{not} efficient. However, compared against the full strong classical simulation schemes described above, the use of a small QPU will (for many instances) prove advantageous as it lessens the demands on the classical machines. This will be discussed in more detail in Section~\ref{subsubsec: PBC - part 2: Hybrid}.

\subsection{The extended Gottesman-Knill theorem} \label{subsec: Extended GK}

We consider (universally general) $n$-qubit unitary Clifford$+T$ quantum circuits with $t$ $T$ gates and $m$ final measurements in the computational basis. Here, the non-Clifford gate $T$ is defined as $T = \mathrm{diag} (1, \, e^{i \pi /4})$. It is well known~\cite{Zhou00, BravKit05} that a $T$ gate can be implemented via the $T$-gadget illustrated in Figure~\ref{fig:T_gadget}, so that the original unitary Clifford$+T$ circuit $\mathcal{U}$ is transformed into an adaptive Clifford circuit $\mathcal{C}$ with input state $\left| 0\right>^{\otimes n} \left| A \right>^{\otimes t}$, $t$ intermediate and $m$ final measurements in the computational basis. Here, the state $\left| A \right>$ is the magic (i.e. non-stabilizer) state defined as $\left| A \right> = ( \left| 0 \right> + e^{i \pi / 4} \left| 1 \right> )/ \sqrt{2}$. These circuits are extremely alluring because they are written exclusively in terms of Clifford gates which can be realized fault-tolerantly using stabilizer codes~\cite{PhDGottesman, Gott98}. However, this implementation comes at the high practical cost of a larger number of logical qubits ($n+t$ instead of $n$). Additionally, it becomes necessary to devise protocols to purify imperfect magic states using Clifford gates only; for a review see Refs.~\cite{CampTV17, Litinski2019}.

In Ref.~\cite{Yoga19}, Yoganathan, Jozsa and Strelchuk used the Pauli-based model of quantum computation introduced in Ref.~\cite{BSS16} (see Section~\ref{subsec: PBC}) to prove that any such adaptive Clifford circuit $\mathcal{C}$ can be reduced (using only a polynomial amount of classical resources) to an adaptive Clifford circuit $\mathcal{C}^{\prime}$ acting only on the magic state input $\left| A \right>^{\otimes t}$ and with at most $t$ measurements in the computational basis. This result--dubbed by the authors the extended Gottesman-Knill theorem (Theorem 3.2 of Ref.~\cite{Yoga19})--clearly provides a conceptual way of compiling general quantum circuits, minimizing the required quantum resources by trading in affordable and efficient (i.e. polynomial) classical processing.

As we will see below, in this context, PBC can be regarded as a useful intermediate step (or proof technique), as in this framework, the reduction from $(n+t)$ to $t$ qubits happens simply and naturally.

\subsection{Pauli-based computation} \label{subsec: PBC}
\subsubsection{Efficient circuit compilation} \label{subsubsec: PBC - part 1}
As mentioned in the last paragraph of the previous Subsection, to understand how any universally general, adaptive Clifford circuit acting on the input state $\left| 0\right>^{\otimes n} \left| A\right>^{\otimes t}$ can be reduced to a quantum computation on only $t$ qubits, it is easier to consider  Pauli-based computation (PBC). Following the terminology used in Ref.~\cite{BSS16}, we define a standard PBC $\mathcal{Q}$ as follows:
\begin{itemize}
    \item The input state is a product state of magic states: $\left| A \right>^{\otimes t}$;
    \item Operations are \textit{non-destructive} measurements of independent and pairwise commuting Pauli operators, $P_i,$ belonging to the Pauli group on $t$ qubits, $\mathcal{P}_t$;
    \item The result of the computation is obtained from the outcomes $s_i$ of the Pauli measurements together with some interleaved (polynomial) classical processing.
\end{itemize}

This definition allows us to write the following theorem (adapted from Ref.~\cite{BSS16}):
\begin{theorem}
Any (universally general) $n$-qubit Clifford+T circuit $\mathcal{U}$ with $t$ $T$ gates can be simulated by a (standard) PBC $\mathcal{Q}$ on $t$ qubits and $\poly(n,t)$ classical processing.
\label{theorem: PBC Universality/ compilation}
\end{theorem}

Theorem \ref{theorem: PBC Universality/ compilation}  is an assertion of the universality of the Pauli-based model of quantum computation but, as we will see, can also be regarded as a statement on the compilation of quantum computations.

The proof of this theorem is quite straightforward. We start by transforming the unitary $n$-qubit Clifford$+T$ quantum circuit into an $(n+t)$-qubit, adaptive Clifford circuit with input state $\left| 0\right>^{\otimes n} \left| A \right>^{\otimes t}$ and $(t+m)$ measurements in the computational basis.

Next, we initialize a generalized PBC list with $n$ dummy single-qubit $Z$ operators acting on each of the main (stabilizer-state) qubits:
\begin{equation}
    \mathrm{(LIST):} \,\,\, Z_1,\, Z_2,\, ...,\,Z_n.
\end{equation}

Following this, we take the single-qubit $Z$ measurement of the first auxiliary qubit, $Z_{a_1}$, and propagate it through all the unitaries to its left until it reaches the very beginning of the quantum circuit. Because the circuit is made up only of Clifford gates, this can be done efficiently~\cite{PhDGottesman, Gott98} and the result will always be a Pauli operator: $Z_{a_1} \rightarrow P_{a_1} = U_{a_1}^{\dagger} Z_{a_1} U_{a_1},$ where $U_{a_1}$ denotes the part of the circuit to the left of the measurement of the first auxiliary qubit and $P_{a_1} \in \mathcal{P}_{n+t}$. Once we have this Pauli operator at the very beginning of the circuit, we proceed in the following manner:

\begin{enumerate}
    \item Assess whether the current Pauli, $P_{a_1}$, commutes or anti-commutes with any of the dummy $Z_i$ operators or with any previously measured and pairwise commuting Pauli operators. (Of course, for the first Pauli, none of the latter operators exist yet, but as they start to arise they need to be considered in this step.)
    
    \item If the current Pauli anti-commutes with a certain Pauli operator $Q$, it is clear its quantum measurement, if performed, would yield the outcome $0$ or $1$ with equal probability. Thence, instead of carrying out an actual (quantum) measurement, we can use a classical computer with a random number generator to randomly draw the measurement outcome $s_{a_1}$ from the uniform distribution $f_U=\{0,1\}$. To ensure that the quantum system is transformed into the correct quantum state that would be obtained after the measurement of $P_{a_1}$, we add the Clifford operator $V(\lambda, s_{a_1} ) = ((-1)^{\lambda} Q + (-1)^{s_{a_1}} P_{a_1})/\sqrt{2}$ at the beginning of the quantum circuit (before all other unitary gates). Here, $\lambda$ denotes the outcome of the previously measured Pauli operator $Q$. Being a Clifford unitary, subsequent Pauli operators will be propagated through $V(\lambda, s_{a_1} )$ easily and efficiently.
    
    \item If, on the other hand, the current Pauli commutes with all previously measured Pauli operators (and dummy $Z$ measurements) there are two possibilities:
    \begin{enumerate}
        \item either it depends on previously measured operators, in which case its outcome $s_{a_1}$ is fully determined by the outcome of those operators, and can be determined classically;
        
        \item or it is independent of previously measured Pauli operators, in which case we need to perform the actual measurement of $P_{a_1} = P_{a_1}^{\prime} \otimes P_{a_1}^{\prime \prime}$, where $P_{a_1}^{\prime} \in \mathcal{P}_{n}$ and $P_{a_1}^{\prime \prime} \in \mathcal{P}_t$ denote the Pauli operators acting on the main (stabilizer-state) and auxiliary (magic-state) qubits, respectively. Note that $P_{a_1}$ is necessarily trivial on the stabilizer-state qubits, i.e. $P_{a_1}^{\prime} = Z^{b_1}\otimes Z^{b_2} \otimes ... \otimes Z^{b_n}$, with $b_i=\{ 0, 1\}$. Therefore, it is clear that we can reduce the measurement of $P_{a_1}$ on all $n+t$ qubits to the measurement of $P_{a_1}^{\prime \prime}$ on the $t$-qubit auxiliary quantum register. 
    \end{enumerate}
\end{enumerate}

\begin{figure*}[t]
    \centering
    \begin{tikzpicture}
     \node[scale=0.8] {
      \begin{quantikz}[thin lines]
        \lstick{\ket{A}} &\gate[wires=2,style={fill=gray!15, rounded corners}]{P_1} &\gate[wires=2,style={fill=gray!15, rounded corners}]{P_2} \\
        \lstick{\ket{A}} &\qw &\qw 
      \end{quantikz}
$\rightarrow$
      \begin{quantikz}[thin lines]
        \lstick{\ket{A}} &\gate[wires=2]{U(P_1)}\gategroup[2,steps=3,style={dashed,rounded corners,fill=blue!20, inner xsep=2pt, inner ysep=10pt},background]{{$P_1$}} &\meter{$s_1$} &\gate[wires=2]{U^{\dagger}(P_1)} &\qw &\gate[wires=2]{U(P_2)}\gategroup[2,steps=3,style={dashed,rounded corners,fill=red!20, inner xsep=2pt, inner ysep=10pt},background]{{$P_2$}} &\qw & \gate[wires=2]{U^{\dagger}(P_2)}\\
        \lstick{\ket{A}} &\qw &\qw &\qw &\qw &\qw &\meter{$s_2$} & 
      \end{quantikz}
     };
    \end{tikzpicture}
    \caption{Implementation of an adaptive PBC via a sequence of adaptively-chosen Clifford circuits, using the approach proposed in Ref.~\cite{Yoga19}. Each colored box corresponds to the implementation of a single Pauli measurement, as described in the main text.}
    \label{fig: PBC implementation by Yoga19}
\end{figure*}
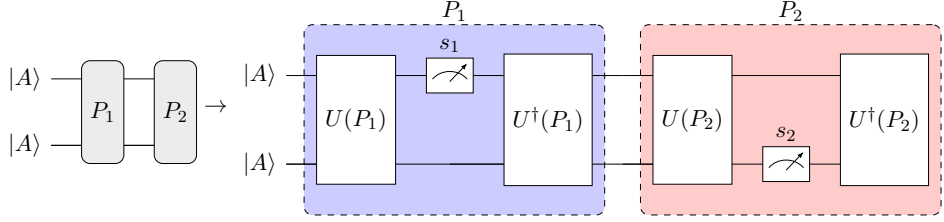

After we have determined the appropriate outcome $s_{a_1}$ for this first Pauli operator $P_{a_1}$, we store it and fix the next part of the adaptive Clifford circuit appropriately. Finally, the Pauli operator is added to the generalized PBC list and (if case 3.(b) happened) to the list of actually measured quantum operators. Then we carry out the procedure described above for each subsequent measurement in the quantum circuit (going first through the measurements of the auxiliary qubits and only afterward through the $m$ measurements of the main qubits), until the outcome of all the Pauli operators in the generalized PBC has either been classically determined (cases 2. and 3.(a)) or actually measured in a given QPU (situation 3.(b)).

The actual outcome of the computation is the bit string $\mathbf{s}$ made up from the $m$ outcomes of the last $m$ Pauli operators (regardless of whether those outcomes were established classically or through real quantum measurements).

The procedure described guarantees that the actual quantum computation corresponds to the measurement of at most $t$ independent and pairwise commuting Pauli operators on the $t$ qubits initially in the product state $\left| A \right> ^{\otimes t}$. Therefore, we are simulating the original circuit by a standard PBC on $t$ qubits, together with some classical processing which takes time $\poly(n,t)$. (For more detailed proofs of some of the claims made along the procedure described above see Refs.~\cite{BSS16} and \cite{Yoga19}.)

It is important to realize that the compiled computation (i.e. the standard PBC) is not always the same. That is, each time we want a single-shot sample from the output distribution of the original unitary Clifford$+T$ circuit, the entire procedure described above needs to be executed, yielding a sequence of Pauli measurements (and, therefore, a standard PBC) which might differ from shot to shot. Only in this way can we guarantee that we are sampling from the correct output distribution.

At this point, we would like to draw the reader's attention to the fact that the extended Gottesman-Knill theorem formulated in Ref.~\cite{Yoga19}  and summarized in the previous Subsection corresponds simply to translating this standard PBC back into the quantum circuit model. In Ref.~\cite{Yoga19}, the authors present a theoretical and abstract way of doing so. According to their proposal, if the outcome of a certain Pauli $P_1 \in \mathcal{P}_t$ needs to be determined via an actual non-destructive quantum measurement (case 3.(b) above), then this can be done via a unitary Clifford circuit $U(P_1)$ which maps the measurement of that $t$-qubit Pauli operator to the measurement of a single qubit (e.g. measurement of $Z_1$). Using this translation back into the quantum circuit model we get the outcome $s_1$ for $P_1$ and can proceed with the PBC procedure described above until we find the next Pauli that needs to be measured, at which point we need to perform another translation back into the circuit model by finding the unitary $U(P_2)$ which maps this new Pauli $P_2$ to the $Z$ measurement of another qubit (e.g. $Z_2$). The general idea is illustrated in Figure~\ref{fig: PBC implementation by Yoga19} for a PBC on $2$ qubits.

In Ref.~\cite{Patel03}, the authors showed that any CNOT circuit of $n$ qubits has an equivalent CNOT circuit with only $O(n^2/ \log n)$ gates. On the other hand, Theorem 8 of Ref.~\cite{AarGott04} states that \textit{``any unitary stabilizer circuit has an equivalent canonical form''} consisting of a sequence of 11 rounds each of which is made up exclusively of either phase, Hadamard or CNOT gates. It is straightforward to see that each phase or Hadamard round requires only $O(n)$ gates, while a CNOT segment can be written with $O(n^2/ \log n)$ gates (as per the result of Ref.~\cite{Patel03} stated above). Returning to the proposal of Ref.~\cite{Yoga19} for the conversion from PBC back to the quantum circuit model, it is easy to understand that each unitary $U(P_i)$ can be efficiently found and implemented via a quantum circuit with $O(t^2/ \log t)$ gates. Since the standard PBC has potentially $t$ independent and pairwise commuting Pauli observables, the $t$-qubit adaptive Clifford circuit proposed in Ref.~\cite{Yoga19} and used to perform the computation has a total number of gates that scales like $O(t^3 / \log t)$. We propose improvements to this in Section~\ref{subsec: Improvement 1: Pauli measurements}.

\subsubsection{Hybrid Pauli-based computation} \label{subsubsec: PBC - part 2: Hybrid}

We have just explained how a certain input quantum circuit can be compiled into an adaptive PBC circuit with $t$ qubits. From this explanation, it is clear that the compilation procedure is naturally interleaved with that of sampling from the output distribution of the original quantum circuit; in fact, the processes of compilation and weak simulation of the original quantum circuit using a QPU with (at least) $t$ qubits together with (polynomial) classical processing are inseparable, and comprise what we name Task 1.

Besides this, Ref.~\cite{BSS16} discusses a related (yet distinct) task, which will now be explained; we call this Task 2. We start by restricting ourselves to the set of input Clifford$+T$ quantum circuits with a single output measurement.\footnote{This could be the case, for instance, of a circuit used to solve a decision problem.} For such input circuits, we want to estimate the probability $p$ of getting the outcome $y=1,$ with an error of (at most) $\epsilon$; in other words, we want to make a strong simulation of the original input circuit. Assuming (just as before) access to a QPU with (at least) $t$ qubits, this task can be achieved simply by carrying out the procedure outlined above a suitable number of times.

The questions posed and answered in Ref.~\cite{BSS16} were the following: (i) can we perform the simulation task described above in a QPU with only $t-k$ qubits?; and (ii) if so, what is the cost of simulating the $k$ virtual qubits?

The first question can obviously be answered in the affirmative. As we have seen, any Clifford$+T$ quantum circuit can be simulated by a standard PBC $\mathcal{Q}$ on $t$ qubits; now, if we only have $t-k$ qubits accessible in our quantum hardware, we can simply decompose the magic state $\left| A \right> \left< A \right|^{\otimes k}$ of the first $k$ qubits into a stabilizer pseudomixture:
\begin{equation}
    \left| A \right> \left< A \right|^{\otimes k} = \sum_{i=1}^{M} \alpha_i \left| \psi_i \right> \left< \psi_i \right| \,,
    \label{eq: |A><A| decomposition}
\end{equation}
where the sum ranges over $M$ pure $k$-qubit stabilizer states $\left| \psi_i \right>\left< \psi_i \right|$ and the coefficients $\alpha_i$ are real numbers. Doing this gives us $M$ generalized PBCs with input state: $\left| \psi_i \right> \left< \psi_i \right| \otimes \left| A \right> \left< A \right|^{\otimes (t-k)}$. Noting that any stabilizer state $\left| \psi_i \right>$ on $k$ qubits can be obtained from the computational basis state $\left| 0 \right>^{\otimes k}$ by the application of a Clifford unitary $U_i$: $\left| \psi_i \right> = U_i \left| 0 \right>^{\otimes k}$, we realize that each of the $M$ generalized PBCs can be transformed into a standard PBC on only $(t-k)$ qubits, following the PBC compilation procedure described above.

As explained in Ref.~\cite{BSS16}, linearity implies that the probability $p$ of getting the output 1 is simply given by:
\begin{equation*}
    p = p_{\mathcal{Q}} = \sum_{i=1}^{M} \alpha_i p_{\mathcal{Q}_i}.
    \label{eq: linearity of the probs}
\end{equation*}
Here, $p_{\mathcal{Q}_i}$ denotes the probability that the output of the $i$-th standard PBC on $(t-k)$ qubits yields 1. 

Now, suppose that $y_i$ is the output of a single shot of the (standard) PBC $\mathcal{Q}_{i}\,.$ It is fairly obvious that $\mathbb{E}(y_i)$ is an unbiased estimator of $p_{\mathcal{Q}_i}$, so that the expectation value $\mathbb{E}(\xi)$ of the random variable $\xi$ defined as:
\begin{equation}
    \xi = \sum_{i=1}^{M} \alpha_i y_i
    \label{eq: probability estimator xi}
\end{equation}
is an unbiased estimator of the desired probability $p$~\cite{BSS16}.

Next, we have to determine the number of samples, $N,$ that need to be drawn from the output distribution of each PBC $\mathcal{Q}_i$ in order to guarantee that $p-\epsilon \leq \mathbb{E}(\xi) \leq p+ \epsilon.$ Note that if we draw $N$ values for each of the random variables $y_i$, this means we have $N$ values for the random variable $\xi$. Thus, the error associated with the estimate of $p$ is given by
\begin{equation*}
    \epsilon = \frac{1}{\sqrt{N}}\sigma (\xi) \leq \frac{1}{\sqrt{N}} \sqrt{\sum_{i=1}^{M} \alpha_i^2}\,.
\end{equation*}

If we want to estimate $p$ with a confidence level of $(1-1/c^2)$, then Chebyshev’s inequality determines that the corresponding confidence interval has the following lower and upper bounds:
\begin{equation*}
    \left[ \mathbb{E}(\xi) - \frac{c \| \alpha \|_2}{\sqrt{N}},\, \mathbb{E}(\xi) + \frac{c \| \alpha \|_2}{\sqrt{N}} \right]\,.
\end{equation*}
where $\| \alpha \|_2 = \sqrt{\sum_{i=1}^{M} \alpha_i^2}$ is the $\ell_2$-norm of the coefficients $\alpha_i$.

Finally, this tells us that if we want a precision $\epsilon$, the number of samples for each PBC needs to be:
\begin{equation}
    N = \Biggl\lceil \left( \frac{c \| \alpha \|_2}{\epsilon} \right )^2 \Biggr\rceil\,.
    \label{eq: Number of samples hybrid}
\end{equation}

Obviously, since there are $M$ different (standard) PBCs involved, the total number of shots is: $\mathcal{N} = M\cdot N$. The overall cost of this procedure is, therefore, directly connected to the value of $M.$ 
For $k=1$ we can write explicitly:

$$
\left| A \right> \left< A \right| = \alpha_1 \left| + \right> \left< + \right| + \alpha_2 \left| - \right> \left< - \right| + \alpha_3 \left| +_i \right> \left< +_i \right| \,,
$$
where $\left| \pm \right>$ are the eigenvectors of the single-qubit Pauli operator $X$ with eigenvalues $\pm 1$, $\left| +_i \right>$ is the eigenvector of the single-qubit Pauli operator $Y$ with eigenvalue $+1$, and the coefficients $\alpha_i$ are
$$
\alpha_1 = \frac{1}{2}\,, \quad \alpha_2 = \frac{1 - \sqrt{2}}{2}\,, \quad \text{and} \quad \alpha_3 = \frac{1}{\sqrt{2}}\,.
$$
Hence, the trivial scaling of $M$ with the number of virtual qubits $k$ is $M = 3^k\,$, giving:
\begin{equation}
    \mathcal{N} = \Biggl\lceil 3^k \left( \frac{c \| \alpha \|_2}{\epsilon} \right )^2 \Biggr\rceil\,.
    \label{eq: Sampling complexity BSS16}
\end{equation}

The ideas described above are the essence of Theorem 3 of Ref.~\cite{BSS16}, stated here in an adapted form:

\begin{theorem}
Any (standard) PBC on $t$ qubits can be simulated by $2^{O(k)}$ (standard) PBCs on $t - k$ qubits and a classical processing which takes time $2^{O(k)} \poly(t,k)\,.$
\label{theorem: hybrid PBC}
\end{theorem}
This Theorem highlights the second big merit of PBC, namely, that it allows a straightforward formulation of a hybrid computation wherein a classical (super)computer can be used to extend the number of qubits available to the QPU (albeit with an exponential overhead).

In Section~\ref{subsec: Improvement 2: Sample complexity}, we will see that the sampling complexity given by Equation~\eqref{eq: Sampling complexity BSS16} can be significantly improved by using a smarter sampling strategy together with a different random variable.

\subsection{Other related works}

The one-way model of measurement-based quantum computation (MBQC)~\cite{RaussBrie2001} has been the focus of significant research over the past two decades. Its proposal by Raussendorf and Briegel in 2001 had disruptive ramifications to the field of quantum computing leading, for instance, to the proposal of all-optical quantum computers~\cite{Nielsen2004, BrowneRudolph2005}, experimental proofs-of-principle of such proposals~\cite{Walther+2005exp1WQC, Prevedel+2007exp1WQC}, and universal blind quantum computation~\cite{BroadFE2009}, amongst others. Since PBC is driven by a sequence of Pauli measurements, it also qualifies as an MBQC model and, similarly to one-way computing, classical feedforward is an essential feature, with the outcome of previous measurements conditioning the ensuing computation. However, there are also points of contrast between the two models. Namely, in one-way computing (magic and stabilizer) single-qubit measurements are performed on an entangled (stabilizer) resource state, successively destroying entanglement as the computation progresses, while in PBC multi-qubit (entangling) Pauli measurements are carried out on a magic state that is initially separable.

Remarkably, contrary to the one-way model, PBC has prompted comparatively little interest in the quantum computing community. One of our goals with this work is to bring PBC to the foreground, evidencing its usefulness in trading (affordable) classical for (expensive) quantum resources. This was partly touched upon in Ref.~\cite{Yoga19}, as mentioned above in Section~\ref{subsec: Extended GK}, where the authors briefly describe how to reduce an adaptive quantum circuit with input $\left| 0 \right>^{\otimes n} \left| A \right>^{\otimes t}$ to a smaller circuit acting only on the $t$ magical qubits, using PBC as a useful tool underlying this reduction. However, because this was not the focus of their work, this observation remained mainly unexplored, with only a broad proposal for the actual implementation of PBC being mentioned briefly therein (see above). Here, we concentrate on this, improving upon their scheme for performing Pauli measurements (Section~\ref{subsec: Improvement 1: Pauli measurements}) and demonstrating the profitability of PBC for circuit compilation and hybrid quantum-classical computation thanks to numerical results obtained with our Python code (Section~\ref{sec: Numerical results}). 

Circuit optimization and compilation are particularly relevant in the current NISQ era but should remain useful in the intermediate and long term as well. Because of this, the literature on these topics is quite extensive with different techniques being explored. Some examples include the use of phase polynomials~\cite{AmyMM2014, NamNSCM2018}, the ZX-calculus~\cite{CowtanDDSS2020, KissWet2020}, and Pauli rotations~\cite{Zhang2019} for optimizing quantum circuits. Here, we introduce PBC into the mix, showing that it can, in many instances, be used to reduce the gate counts and depth needed for the computation but also the number of qubits. However, PBC compilation is fundamentally different from any of the aforementioned compilation techniques in that it is inextricable from the task of simulating the original quantum circuit. That is, with PBC, we get an all-in-one compiler and (weak) simulator. This is unlike other compiling tools that take as input a certain quantum circuit and output an optimized version of it that can be run separately \textit{a posteriori}. 

On a different note, the hybrid computation scheme for simulating a large quantum circuit using a smaller one (Sections~\ref{subsubsec: PBC - part 2: Hybrid} and \ref{subsec: Improvement 2: Sample complexity}) is akin to circuit knitting techniques~\cite{BSS16, PengHarrowOW2020, Tang+2021, Piveteau2023, Lowe+2023}, which have seen a rich development in recent years. Broadly speaking, these techniques divide into two main categories: (i) gate (space-like) cutting, and (ii) wire (time-like) cutting. Regardless of the approach, the essential point is that these knitting techniques cut a large quantum circuit into multiple smaller quantum circuits that can hopefully be run in near- or intermediate-term devices, where the number of qubits is still limited. The drawback is that these techniques are restricted to the estimation of expectation values of observables via a sampling procedure whose number of samples is exponential in the number of cuts performed. In this regard, we see similarities with the hybrid PBC procedure, where the sampling complexity also scales exponentially, this time with the number of virtual qubits. The main difference between hybrid PBC and circuit knitting is that in the former we remove a desired number of virtual qubits which are essentially dealt with classically, while in the latter the computation is broken into two (or more) smaller quantum computations.

Pauli measurements take a prominent role in PBC, as they constitute the computational steps for this model. They are also a crucial component of quantum error correction, where syndrome measurements allow for the detection (and posterior correction) of errors~\cite{Gottesman2009, FowlerMMC2012, Litinski2019GoSC}. Particularly, in Ref.~\cite{Litinski2019GoSC}, Litinski discusses how to run quantum circuits in a fault-tolerant architecture with minimal overhead, focusing on strategies for surface-code quantum computing. Some similarities can be identified between that work and the PBC compilation technique explored herein. Namely, the transformation of any quantum circuit into one described only in terms of non-stabilizer (multi-qubit) Pauli rotations followed by (multi-qubit) Pauli measurements is very close in spirit to the PBC procedure proposed in Ref.~\cite{BSS16} and summarized in Section~\ref{subsec: PBC}. Nevertheless, Litinski uses these techniques in the context of setting up a fault-tolerant quantum computer, discussing magic state distillation and computation with logical qubits. In contrast, all of our discussions happen at the physical level, as we are mostly concerned with quantum computing in the near- and intermediate-term, before fault tolerance becomes available.

\section{Optimized implementations of Pauli-based computation} \label{sec: Main contributions}

\subsection{Pauli measurements}\label{subsec: Improvement 1: Pauli measurements}
\begin{figure*}[t]
    \centering
    \begin{tikzpicture}    \node[scale=0.8] {       \begin{quantikz}[thin lines]         \lstick{\ket{A}} &\gate[wires=2,style={fill=gray!15, rounded corners}]{P_1} &\gate[wires=2,style={fill=gray!15, rounded corners}]{P_2} \\         \lstick{\ket{A}} &\qw                                                       &\qw        \end{quantikz}
$\rightarrow$
      \begin{quantikz}[thin lines]                 \lstick[wires=2]{$\ket{A}^{\otimes 2}$} &\qw\gategroup[wires=3,steps=5,style={dashed,rounded corners,fill=blue!20, inner xsep=2pt},background]{{$P_{1} = \sigma_{1;1} \otimes \sigma_{1;2}$}} &\gate{\sigma_{1;1}} &\qw            &\qw      &\qw                &\qw                                                        &\qw\gategroup[wires=3,steps=5,style={dashed,rounded corners,fill=red!20, inner xsep=2pt},background]{{$P_{2} = \sigma_{2;1} \otimes \sigma_{2;2}$}}  &\gate{\sigma_{2;1}} &\qw            &\qw      &\qw &\qw\rstick[wires=3]{\ket{\Psi_f}} \\                                                         &\qw                                                                                                                                        &\qw            &\gate{\sigma_{1;2}} &\qw      &\qw                &\qw                                                        &\qw                                                                                                                                        &\qw            &\gate{\sigma_{2;2}} &\qw      &\qw &\qw                               \\                 \lstick{$\ket{0}_{aux}$}                &\gate{H}                                                                                                                                   &\ctrl{-2}      &\ctrl{-1}      &\gate{H} &\meter{$s_1$} &\gate[style={fill=black},label style=white]{\ket{0}_{aux}} &\gate{H}                                                                                                                                   &\ctrl{-2}      &\ctrl{-1}      &\gate{H} &\meter{$s_2$}       \end{quantikz}    }; \end{tikzpicture}
    \caption{Our first proposal for carrying out the sequence of non-destructive measurements in the adaptive PBC obtained during the compilation process. Each colored box corresponds to the implementation of a single Pauli measurement, as described in the main text. The black box with label $\left| 0\right>_{aux}$ indicates a re-setting of the auxiliary qubit in state $\left| 0\right>$.}
    \label{fig: Our PBC implementation}
\end{figure*}
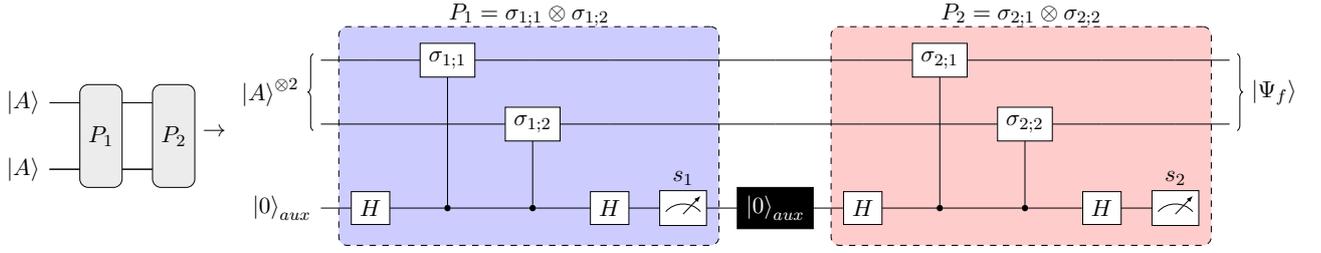

In this Section, we introduce three alternative ways of implementing a standard PBC, i.e. of carrying out a sequence of $t$ adaptive and non-destructive measurements of independent and pairwise commuting Pauli operators. These schemes can be compared to the prior work by Yoganathan \textit{et al.}~\cite{Yoga19} presented above.

\subsubsection{$O(t^2)$ depth with one auxiliary qubit}\label{subsubsec: Proposal 1}

Our first approach is inspired by quantum-error correction and syndrome measurement of Pauli operators~\cite{Gottesman2009} and requires the use of one auxiliary qubit whose measurement determines the outcome of the corresponding $t$-qubit Pauli operator. Consider that the observable to be measured (non-destructively) is $P_i = \sigma_{i;1} \otimes \sigma_{i;2} \otimes ... \otimes \sigma_{i;t}$, where each $\sigma_{i;j}$ denotes a single-qubit Pauli operator $I$, $X$, $Y$, or $Z$. Our method works by bringing in an auxiliary qubit in the state $\left| + \right> = H \left| 0 \right>$, and using it as the control qubit for the sequence of controlled-$\sigma_{i;j}$ operations each of which targets the corresponding main qubit $j$. After that sequence, a Hadamard gate is applied to the auxiliary qubit which is then measured in the computational basis. It is easy to show that, after this measurement, the main qubits are left in a state which is the $(-1)^{s_i}$-eigenstate of $P_i$, where $s_i \in \{0,1\}$ is the outcome of the measurement of the auxiliary qubit. This qubit can be reset and re-used to measure the next Pauli in the (adaptive) PBC sequence.\footnote{Note that resetting can be easily carried out by simply adding an $X$ gate whenever necessary, i.e. whenever the outcome is 1.} This proposal is illustrated in Figure~\ref{fig: Our PBC implementation} for the case of a simple $2$-qubit PBC. 

Clearly, the implementation of a $t$-qubit PBC using this scheme has a total depth of $O(t^2)$ layers; additionally, the total number of gates also scales like $O(t^2)$, which is a significant improvement over the proposal in Ref.~\cite{Yoga19}.

We can consider this analysis a bit more carefully and determine upper bounds for the total number of single- and $2$-qubit Clifford gates, and depth of the overall (PBC-compiled) quantum circuits within this framework. We construct the compiled circuits using only the Clifford generators: $H$, $S$ and CNOT. This is not a problem as the controlled-$Y$ and controlled-$Z$ operations used in our procedure can be easily written in terms of this set of gates using the identities: $CY_{ij} = S_j CX_{ij} S_j^3$ and $CZ_{ij} = H_j CX_{ij} H_j.$ The maximum number of operations and depth arise in the case where each of the $t$-qubit Pauli operators has $(t-1)$ single-qubit $Y$ operators and exactly $1$ Pauli $Z$ operator, for instance: $P_1=Z\otimes Y\otimes Y\otimes ... \otimes Y,$ $P_2=Y\otimes Z\otimes Y\otimes ... \otimes Y,$ etc. These operators are all independent and pairwise commuting and give rise to an upper bound on the number of single-qubit Clifford generators which is $N_{HS}^{ub} = 4t^2$ and an upper bound on the total number of CNOT gates given by $N_{\mathrm{CNOT}}^{ub} = t^2.$ As for the depth, the upper bound within this scheme is $D^{ub} = t(t+5) - 1$ layers. Evidently, these upper bounds will often be gross over-estimates of the actual depth and gate counts of the PBC-compiled circuits as the Pauli observables are commonly trivial in several of the qubits. This will be observed in the numerical results presented in Section~\ref{sec: Numerical results}. Additionally, in that Section, we will also confirm that this implementation of the PBC-compiled circuits frequently leads to reductions in the depth and gate counts needed to perform the computation. Moreover, we see that the number of qubits of the PBC-compiled circuits is reduced with respect to that of the original Clifford$+T$ quantum circuit whenever $t< n - 1\,.$

\subsubsection{$O(t \log t)$ depth without auxiliary qubits}\label{subsubsec: 2nd proposed scheme -- t*logt depth}
\begin{figure*}[t]
    \centering
    \begin{tikzpicture}
   \node[scale=0.8]{
      \begin{quantikz}[thin lines]
         \lstick[wires=4]{$\ket{\Psi_{in}}$}  &\gate[wires=4,style={fill=gray!15, rounded corners}]{P = \sigma_1 \otimes \sigma_2 \otimes \sigma_3 \otimes \sigma_4}  &\qw\rstick[wires=4]{\ket{\Psi_f}} \\
                                              &\qw  &\qw \\
                                              &\qw  &\qw \\
                                              &\qw  &\qw 
      \end{quantikz}

$\iff$

      \begin{quantikz}[thin lines]
                \lstick[wires=4]{$\ket{\Psi_{in}}$}  &\gate{C_1}\gategroup[wires=4,steps=3,style={dashed,rounded corners,fill=purple!20, inner xsep=2pt},background]{}  &\ctrl{1}  &\qw       &\qw          &\qw\gategroup[wires=4,steps=3,style={dashed,rounded corners,fill=teal!20, inner xsep=2pt},background]{{uncomputing}}  &\ctrl{1}  &\gate{C^{\dagger}_1}  &\qw\rstick[wires=4]{\ket{\Psi_f}} \\
                                                     &\gate{C_2}                                                                                                        &\targ{}   &\ctrl{2}  &\qw          &\ctrl{2}                                                                                                              &\targ{}   &\gate{C^{\dagger}_2}  &\qw \\
                                                     &\gate{C_3}                                                                                                        &\ctrl{1}  &\qw       &\qw          &\qw                                                                                                                   &\ctrl{1}  &\gate{C^{\dagger}_3}  &\qw \\
                                                     &\gate{C_3}                                                                                                        &\targ{}   &\targ{}   &\meter{$s$}  &\targ{}                                                                                                               &\targ{}   &\gate{C^{\dagger}_4}   &\qw
      \end{quantikz}
   };
\end{tikzpicture}
    \caption{Our second proposal for performing non-destructive measurements of multi-qubit Pauli operators. In this case, the arbitrary $4$-qubit Pauli operator $P=\sigma_1\otimes \sigma_2\otimes \sigma_3\otimes \sigma_4$ is to be measured on an arbitrary state $\left| \Psi_{in} \right>.$ This can be done via the quantum circuit depicted on the right, which uses a logarithmic-depth cascade of CNOTs. The gates $C_i$ denote the single-qubit Clifford gates such that $\sigma_i = C_i^{\dagger} Z C_i\,.$ In the block highlighted in teal we \textit{uncompute} the CNOTs and single-qubit Clifford transformations; this is necessary to ensure that the correct $(-1)^s$-eigenstate of $P$, $\left| \Psi_{f}\right>$, is obtained at the end.}
    \label{fig: 2nd implementation}
\end{figure*}
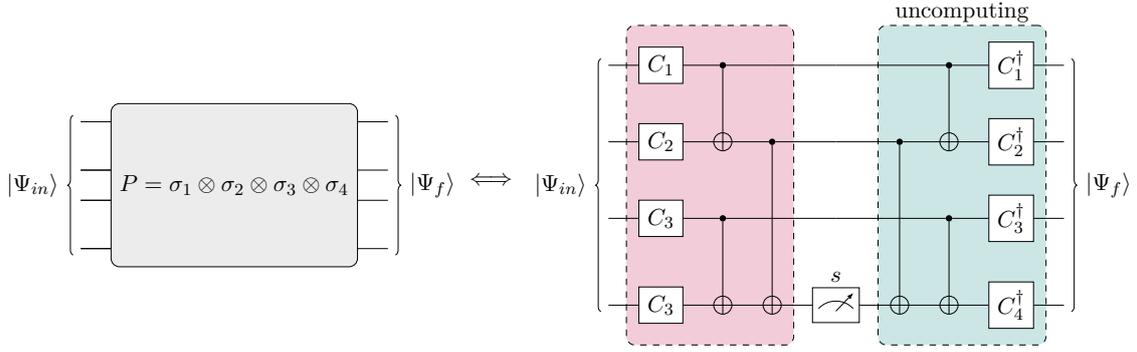

It is known that logarithmic-depth quantum circuits can be used for computing $n$-bit parity functions in quantum circuits; see, for instance, Figure~3 of Ref.~\cite{ChoiMeter2011}. By making some adjustments, this kind of construction can be adapted to our task of measuring arbitrary $t$-qubit Pauli operators, $P=\sigma_1\otimes \sigma_2 \otimes \dots \otimes \sigma_t,$ and leads to an improvement of the depth from $O(t^2)$ to $O(t \log t)$ while avoiding the additional auxiliary qubit of the previous construction.

A depiction of the scheme can be seen in Figure~\ref{fig: 2nd implementation} for the case of an arbitrary 4-qubit Pauli operator, but it generalizes to any $t$-qubit Pauli measurement. First, as seen in the Figure, we start by choosing one of the data qubits as the qubit to be measured, in the present case, qubit 4. We apply the cascade of CNOTs analogous to that of Ref.~\cite{ChoiMeter2011}, but before that, we need to add appropriate single-qubit Clifford unitaries $C_i$ such that $\sigma_i = C_i^{\dagger} Z C_i\,.$ After the fourth qubit has been measured, yielding an outcome $s$, it is necessary to \textit{uncompute} the cascade of CNOTs and single-qubit Cliffords to ensure that the system is left in the correct $(-1)^s$-eigenstate of $P$. 

A fairly simple way of seeing that this construction works properly is to backward propagate the single-qubit measurement performed on qubit 4 until it reaches the beginning of the quantum circuit. Because the two highlighted circuit blocks are transposed conjugates of each other, it is clear that they cancel out after the measurement has been pushed to the beginning and transformed into the $4$-qubit Pauli measurement $P = C_1^{\dagger}ZC_1 \otimes C_2^{\dagger}ZC_2 \otimes C_3^{\dagger}ZC_3 \otimes C_4^{\dagger}ZC_4\,.$

It is easy to see that when $\sigma_i = Z$ then $C_i$ corresponds to the identity, whilst if $\sigma_i = X$ or $\sigma_i = Y$ we have $C_i = H $ and $C_i = H S^\dagger $, respectively. Additionally, we note that the construction in Figure~\ref{fig: 2nd implementation} assumes that the Pauli operator is non-trivial in all qubits. If that is not the case, a logarithmic-depth cascade is still possible, with small modifications. For instance, if $P=I\otimes \sigma_2 \otimes \sigma_3 \otimes \sigma_4$, we need only remove the CNOT gates between qubits 1 and 2. Analogously, if $P=\sigma_1 \otimes \sigma_2 \otimes I \otimes \sigma_4,$ the CNOT gates between qubits 3 and 4 should be removed. A trickier case arises if $P=\sigma_1 \otimes I \otimes \sigma_3 \otimes \sigma_4,$ wherein we cannot simply remove the controlled-\textsc{not} gate between the second and the fourth qubits without further changes. In fact, in order to implement such an example, we need to remove that gate and change $CX_{1,2}$ to $CX_{1,4}.$ Finally, if the Pauli measurement is trivial on the fourth qubit, then the setup should be changed so that a different qubit is chosen to be measured, for instance, qubit 1. This is because whichever qubit holds the computational basis measurement needs to correspond to a qubit wherein the Pauli operator acts non-trivially. In any case, provided one is careful with these details, it is always possible to guarantee a logarithmic-depth quantum circuit for an arbitrary $t$-qubit Pauli measurement.

Since each Pauli can now be measured in $O(\log t)$ depth, this leads to the promised depth improvement of PBC-compiled circuits from $O(t^2)$ to $O(t \log t)$ while maintaining the use of $O(t^2)$ gates but avoiding the need for an additional auxiliary qubit.

As a final remark, we note that we can go even further in terms of saving quantum resources as the ``\textit{uncomputing}'' block, depicted in teal in Figure~\ref{fig: 2nd implementation}, can be dispensed with from the actual quantum computation. This is done by recognizing that the effect of that block in the whole PBC procedure will be to transform upcoming Pauli operators into different ones under conjugation. Therefore, after performing the circuit slice in the magenta box and measuring, we can store the information regarding the uncomputing block and use it only within the classical side-processing to update the ensuing Pauli observables appropriately but never in the actual quantum hardware.

\subsubsection{$O(t)$ depth with $t$ auxiliary qubits}\label{subsubsec: 3rd proposed scheme -- linear depth}

To conclude, a trade-off can be made between the depth of the PBC-compiled quantum circuits and the number of auxiliary qubits. In particular, we point out that it is possible to reduce the depth of the computation from the previous $O(t \log t)$ to $O(t)$ at the expense of requiring $t$ auxiliary qubits rather than none. In this case, the number of qubits of the PBC-compiled circuits is only lower than that of the original circuit if $2t<n$\,. Nevertheless, even if this condition is not met, reducing the depth is a worthy goal in itself as it reduces the demands on the coherence time of the qubits. Therefore, deciding which construction is better suited to practical implementation needs to be done on a case-by-case basis.

Before proceeding, we would like to point out that this proposal has recently been put forth in a paper generalizing PBC for higher-dimensional qudits~\cite{Peres2023}. Nevertheless, it applies to the qubit case just the same.

To decrease the depth of the compiled (adaptive) quantum computation from $O(t \log t)$ to $O(t)$, we apply the following procedure in order to measure a $t$-qubit Pauli operator $P = \sigma_{1}\otimes \sigma_{2}\otimes \dots \otimes \sigma_{t}$:
\begin{enumerate}
    \item Initialize the $t$ auxiliary qubits in the GHZ state $\left| \mathrm{GHZ}_t \right> = \left( \left| 0 \right>^{\otimes t} + \left| 1 \right>^{\otimes t} \right)/\sqrt{2}.$ Such state can be generated by a quantum circuit of constant depth as depicted in Figure~1 of Ref.~\cite{Quek+2022}.
    
    \item For each $\sigma_{i}$ in $P$ apply a controlled-$\sigma_{i}$ operation from the auxiliary qubit $i$ to the main qubit $i$. Clearly, this can now be done in a single layer.
    
    \item Next, apply a Hadamard gate to each of the qubits of the GHZ state and measure them in the computational basis to get a sequence of outcomes $\{s_i\}_{i=1}^t$. This leaves the main qubits in the proper $(-1)^{s}$-eigenstate of $P$ operator, where $s = \bigoplus_{i=1}^t s_i$ is the measurement outcome.
\end{enumerate}
It is clear that the procedure described above allows for the measurement of any $t$-qubit Pauli observable via a circuit of constant depth. Since there are at most $t$ such operators in the PBC sequence, the compiled circuit will have a depth that scales as $O(t),$ as promised.

Note that the total number of operations in this approach still scales as $O(t^2)$, but the use of the $t$-qubit GHZ state allows the controlled-$\sigma_{i}$ operations in each Pauli to be carried out simultaneously (in a single layer) making the circuit shallower and significantly reducing the demands on the coherence time of the qubits.

\vspace{0.5cm}\noindent\textbf{Remark.} As a final comment, we would like to point out once more that PBC compilation may significantly reduce the depth and gate count of a given input circuit. This is because the compiled circuits have a standard form, corresponding to at most $t$ Pauli measurements. As we have seen, this standard form has strict and easy-to-derive bounds on gate count and depth, and which do not depend on the number of qubits, depth, or Clifford gate count of the original circuit, but only on the number of $T$ gates. This feature will be clear in the numerical results presented in Section~\ref{sec: Numerical results}.

\subsection{Sampling complexity of hybrid computation}\label{subsec: Improvement 2: Sample complexity}
\begin{table*}[]
    \centering
    \begin{tabular}{ c||c|c||c|c||c|c||}
            & \multicolumn{2}{|c||}{Prior work~\cite{BSS16}}& \multicolumn{2}{|c||}{In our code} & \multicolumn{2}{|c||}{Best to date} \\
        \hline\hline
        $k$ & $\epsilon=0.1$ & $\epsilon=0.01$ & $\epsilon=0.1$ & $\epsilon=0.01$ & $\epsilon=0.1$ & $\epsilon=0.01$ \\
        \hline
        1 & $23\,790$ & $2\,378\,685$ & $530$ &   $52\,984$    & $530$ & $52\,984$ \\
        2 & $56\,610$ & $5\,658\,120$ & $1\,060$ &   $105\,967$   & $810$ & $80\,904$ \\
        3 & $134\,595$ & $13\,458\,960$ &  $2\,120$   &   $211\,933$   & $1\,305$  & $130\,438$ \\
        4 & $320\,355$ & $32\,014\,440$ &  $4\,239$   &   $423\,866$   & $2\,172$  & $217\,107$ \\
    \end{tabular}
    \caption{Required number of samples, $\mathcal{N}$, as a function of the number of virtual qubits, $k$, of the desired maximum additive error, $\epsilon$, and for three different sampling strategies. The values were computed considering a 99\% confidence level so that the coefficient $c$ in Equation~\eqref{eq: Sampling complexity BSS16} is $c=10 $ and the probability of failure in Equation~\eqref{eq: Improved sampling complexity hybrid} is $p_{fail}=0.01$. The first two columns yield the number of samples obtained from Equation~\eqref{eq: Sampling complexity BSS16}, following the work of Bravyi, Smith and Smolin~\cite{BSS16}. The two middle columns show the values for that same quantity when using our code. Finally, the last two columns present the best sampling values, obtained using the decompositions of $\left| A \right> \left< A \right| ^{\otimes k}$ which minimize the $\ell_1$-norm, $\| \alpha\|_1$, of the coefficients $\alpha_i$, as computed in Ref.~\cite{Heinrich2019}.}
    \label{tab: Hybrid_PBC-nr_samples}
\end{table*}

Our second improvement consists of reducing the sampling complexity presented in Equation~\eqref{eq: Sampling complexity BSS16}. We can recognize that the approach described in Section~\ref{subsubsec: PBC - part 2: Hybrid} is a bit naive in that the $3^k$ terms in the decomposition in Equation~\eqref{eq: |A><A| decomposition} are all sampled from equally.

Hence, one way of reducing the number of samples required to achieve an estimate of $p$ with precision $\epsilon$ and probability of failure $p_{fail}$ (or, equivalently, confidence level $1-p_{fail}$) is to use a sampling strategy akin to the ones used in quasi-probability-based methods of classical simulation (e.g.~\cite{Pashayan15, HowCamp17}). More specifically, we use the coefficients $\alpha_i$ to define a (true) probability distribution
$\pi_i = \left| \alpha_i \right| / \| \alpha\|_1\,$
where $\| \alpha\|_1 = \sum_{i=1}^M \left|\alpha_i \right|$ is the $\ell_1$-norm of the coefficients $\alpha_i$. Next, we sample an $i$ value from the probability distribution $\pi_i$ and run the corresponding PBC (on $t-k$ qubits) to get the outcome $y=\{0,1 \}$.

It is straightforward to show~\cite{Peres2023,HowCamp17} that the random variable
$$
\eta = \frac{1}{2} - \frac{1}{2}\mathrm{sign}\left(\alpha_i\right) (-1)^y \| \alpha \|_1 
$$
is an unbiased estimator of the probability $p$. Here, $\mathrm{sign}\left(\alpha_i\right)$ denotes the signal of the sampled coefficient $\alpha_i$. Since the random variable is bounded, $\eta \in \left[ \left(1 - \| \alpha \|_1 \right) / 2,\, \left(1 + \| \alpha \|_1 \right) / 2 \right],$ Hoeffding's inequality guarantees that the number of samples needed to estimate $p$ using this sampling strategy is
\begin{equation}
    \mathcal{N} = \Biggl\lceil \frac{\|\alpha\|^2_1}{2 \epsilon^2} \ln\left( \frac{2}{p_{fail}}\right) \Biggr\rceil\,.
    \label{eq: Improved sampling complexity hybrid}
\end{equation}

In this case, the exponential scaling is ``hidden away'' in the $\ell_1$-norm of the coefficients. The minimum value assumed by this quantity constitutes a magic monotone called \textit{robustness of magic} (RoM), $\mathcal{R}$~\cite{HowCamp17, Heinrich2019}. Clearly, optimal decompositions, that is decompositions with minimal $\|\alpha\|_1$ achieve the best possible sampling complexity. Particularly, the results of Ref.~\cite{Heinrich2019} allow us to state that $\mathcal{N} = O(2^{0.7374k} \epsilon^{-2})$ at best.

For simplicity of implementation, our code does not consider the best (known) decompositions obtained in Ref.~\cite{Heinrich2019}. Instead, it uses the fact that the RoM for a single copy of the magic state $\left| A \right>$ is $\mathcal{R}_1 = \sqrt{2}$ and, for larger $k$, simply takes tensor products of that decomposition; thus, the scaling goes with $\|\alpha\|^2_1 = 2^{k}$ which is strictly greater than $\mathcal{R}_k^2$ for any $k>1.$ As a consequence, our code does not run with the best possible sampling complexity but, even still, the number of samples given by Equation~\eqref{eq: Improved sampling complexity hybrid} scales very favorably when compared with the naive result in Equation~\eqref{eq: Sampling complexity BSS16}. Some numerical values can be seen in Table~\ref{tab: Hybrid_PBC-nr_samples} for a confidence level of 99\% and precision of either $\epsilon=0.1$ or $\epsilon=0.01$. 

The fact that the sampling complexity is directly related to the RoM allows us to establish not only upper but also lower bounds for this quantity. Specifically, for any $n$-qubit state $\left| \psi \right>,$ it has been shown~\cite{LeoneOH2022} that the RoM is related to the $\nicefrac{1}{2}$-stabilizer Rényi entropy, $M_{\nicefrac{1}{2}}\left( \left| \psi \right> \right)$, by $M_{\nicefrac{1}{2}} \left( \left| \psi \right> \right) \leq 2 \log_2 \mathcal{R}\left( \left| \psi \right> \right),$ with
\begin{equation*}
    M_{\nicefrac{1}{2}} \left( \left| \psi \right> \right) = 2 \log_2 \left( 2^{-n} \sum_{P\in \mathcal{P}_n} \left| \left< \psi \right| P\left| \psi \right> \right| \right)\,.
\end{equation*}
Stabilizer Rényi entropies are additive, so that for the product state $\left| A \right>^{\otimes k}$ we have~\cite{Peres2023,LeoneOH2022}
\begin{equation*}
    M_{\nicefrac{1}{2}} \left( \left| A \right>^{\otimes k} \right) = M_{\nicefrac{1}{2}} \left( \left| A \right> \right) k = 2 \log_2 \left( \nicefrac{(\sqrt{2} + 1)}{2}\right)k\,,
\end{equation*}
which leads to a sampling complexity lower bound of $\mathcal{N} = \Omega (2^{0.5431k} \epsilon^{-2})\,.$

\section{Practical contributions and numerical results} \label{sec: Numerical results}

\subsection{Efficiency and complexity} \label{subsec: Complexity}
Besides the theoretical proposals presented in the previous Section, we wrote Python code that actually performs the compilation of Clifford$+T$ circuits with $t$ T gates, reducing the quantum computation to a standard PBC on $t$ qubits, assisted by polynomial classical processing (as described in Section~\ref{subsec: PBC}). This code can be found at \url{https://github.com/fcrperes/CompHybPBC}, together with some helpful documentation and toy examples. We defer any specific details of our code implementation to that repository. Here, it is only important to point out that the input quantum circuit must be unitary and written in terms of the Clifford generators $H$, $S$, and CNOT, together with the $T$ gate. Moreover, in alignment with the discussion above, the final quantum computation will be implemented using only these Clifford generators, and the first of our three proposed schemes. 

Unfortunately, the possibility of running quantum circuits incorporating classical computation interleaved with quantum operations is still a bit restricted. This left us facing a technological limitation as we were unable to actually implement the compiled, adaptive computation in actual quantum hardware. However, a reader with access to such technology can easily adapt our code to carry out our proposed simulation scheme. Recently, \textit{dynamic} quantum circuits have become publicly available~\cite{IBMdynamic}, so that our code can now, in principle, be easily used with these quantum chips by anyone who would like to further explore this tool.

At present, however, and to circumvent the aforementioned limitation, we settled for a simple proof-of-principle, resorting to Qiskit's Statevector Simulator~\cite{IBMstatevec}, a Schr{\"o}dinger-type simulator that we use to store and update the wavefunction after each circuit slice. This creates memory and runtime limitations for this proof-of-principle, but those constraints are removed (albeit replaced by others) if dynamic quantum/classical computation is used instead, requiring only the rewriting of a single function in our code.

Having the future deployment of these techniques in real quantum hardware in mind, it is clear that the feasibility of PBC compilation demands that it be carried out within the coherence time of the qubits. Thus, in the next Subsection, we will assess the total runtime of our code, dividing it into what we call the classical and quantum runtimes. The former comprises the time it takes to carry out all the (polynomial) classical side-processing in a single shot of the PBC compilation procedure, while the latter refers to the time it takes to execute the actual quantum job (also per shot). Because, as explained above, our prototype is not integrated with a real QPU, we focus our analysis on the classical runtime of our code.

We have claimed multiple times that the classical processing carried out during the PBC compilation procedure takes polynomial time. We will now discuss this in a bit more detail, for the particular case of our implementation. As we have seen, the first step of the compilation procedure consists of propagating a single-qubit $Z$ measurement to the beginning of the quantum circuit; this task is accomplished in our code in $O((n+t) (m+t) + g + t)$ time, where $m$ and $g$ are, respectively, the number of measurements and Clifford gates in the original Clifford$+T$ circuit. Once we have the $(n+t)$-qubit Pauli operator at the beginning of the circuit we evaluate whether it commutes or anti-commutes with any operator of the list of previously measured, independent and pairwise commuting Pauli operators. In our implementation, this is done in $O(t(n+t))$ time in the worst case and in $O(n)$ time at best. As we have seen, if the Pauli observable indeed anti-commutes with any previously measured operator, we need only make a coin toss to ascertain its outcome, compute the Clifford operator $V$, and add it to the circuit. This is done in time $O(n + g + t + m)$; in doing this, the assessment of that Pauli observable is completed, allowing us to move on to the next one. However, if the Pauli commutes with all previously measured Pauli operators, we have seen that it is necessary to verify if it is dependent on or independent of those Pauli operators. Moreover, if it depends on a (sub)set of the previously performed (independent and pairwise commuting) Pauli measurements, it is necessary to identify precisely what that (sub)set is, so as to classically ascertain the (deterministic) outcome of the current Pauli. Carrying out this task has a cost that scales like $O(t^2 + (n + t))$ in the worst case.

Considering the entire PBC compilation procedure, the worst-case cost of updating a single Pauli operator is $O((n+t)(m+t) + t^2 + g)$. Since there is a total of $(m+t)$ Pauli operators that need to be propagated through the circuit and assessed at the end, the total cost of the computation is, in the worst case,
\begin{equation*}
    O\left((m+t) [(n+t)(m+t) + g]\right).
\end{equation*}
Assuming the situation where $m=n,$ this simplifies further to:
\begin{equation}
    O\left(n^3 + n^2t + nt^2 + t^3 + g(n+t)\right).
    \label{eq: Big-O complexity}
\end{equation}
This Equation gives a behavior for the classical runtime that is cubic in both $n$ and $t$ and linear in $g$. However, it is important to point out that the costs presented are the worst-case costs, computed assuming the largest possible size of the matrix that stores in its rows the independent and pairwise commuting Pauli operators measured up until a given point in the PBC procedure. It is clear that the size of this matrix grows as the computation progresses (i.e. as we find new independent and pairwise commuting Pauli operators) so that earlier steps are much cheaper than estimated and, thus, the amortized cost of the computation is lower than the results presented above. This is clear from the numerical results for the time of the classical processing presented below (cf. Figures~\ref{fig: HSCs--classical_runtime} and \ref{fig: RQCs--classical_runtime}).

The discussion presented above applies to the PBC compilation procedure together with the task of weak simulation. From Sections~\ref{subsubsec: PBC - part 2: Hybrid} and \ref{subsec: Improvement 2: Sample complexity}, it is clear that things change slightly for the task of hybrid PBC. Evidently, the compilation procedure is still an integral part of that task, retaining its polynomial efficiency and cost (but with $m=1$). However, as seen above, we now have a scaling given by $O\left( 2^{0.7374k} \right) \poly(t,k)$ at best. Therefore, it is clear that there is an exponential demand on the side of the quantum hardware and also, inevitably, on the overall classical processing. Given this exponential overhead, it is fair to ask ourselves whether this approach brings any advantage over the full classical simulators discussed in Section~\ref{subsec: Classical simulation of quantum circuits}. Considering those based on stabilizer-rank techniques, we have seen that the best known scaling of such simulators for the task of strong simulation is $O(2^{0.3693t}).$ Considering this asymptotic behavior and comparing it against the one achieved with hybrid PBC, we see that our approach is asymptotically advantageous whenever $k \lesssim t/2\,.$\footnote{Evidently, this comparison is a bit over-simplistic in the sense that it disregards important considerations of practical nature, such as, for instance, the presence of noise in a real quantum device.}

To benchmark our code we run it on a remote server with a total available memory of 32 GB and 8 Intel(R) Xeon(R) CPU E3-1270 v5 @ 3:60 GHz processors each with 4 CPU cores. We start by considering the problem of sampling from the output distribution of the given input quantum circuits, and afterward we consider the task of carrying out approximate strong simulation with different additive errors and different numbers of virtual qubits. For these two distinct tasks we benchmark our simulations by considering first hidden-shift circuits (HSCs), followed by random quantum circuits (RQCs).

We would like to conclude by pointing out that the main goal of our code is to materialize the abstract and conceptual ideas of PBC into a practical solution for compiling Clifford$+T$ quantum circuits. Therefore, the runtime of our code can certainly be significantly improvable via more sophisticated software-engineering techniques, not to mention the use of a different programming language like C++ or Julia.

\subsection{Task 1: Compilation and weak simulation}\label{subsec: Results - task 1}

To analyze the performance of our compilation code, each input Clifford$+T$ unitary quantum circuit is compiled and simulated for a total of $1\,024$ shots, and we evaluate the classical processing time per shot, as well as the total single- and $2$-qubit gate counts, and depth obtained for each compiled circuit.

\subsubsection{Hidden-shift circuits} \label{subsubsec: HSA}

Hidden-shift circuits (HSCs) have been frequently used in the literature to benchmark the performance of full classical simulators using stabilizer-rank-based techniques~\cite{BG16, BBCCGH19, KissWet21}. This is because, on the one hand, they are naturally written in terms of the Clifford$+T$ gate set and, on the other hand, they are deterministic, so that the outcome of the simulator can be verified against the correct outcome (known \textit{a priori}).

\begin{figure}[t]
    \centering
    \includegraphics[scale=0.5]{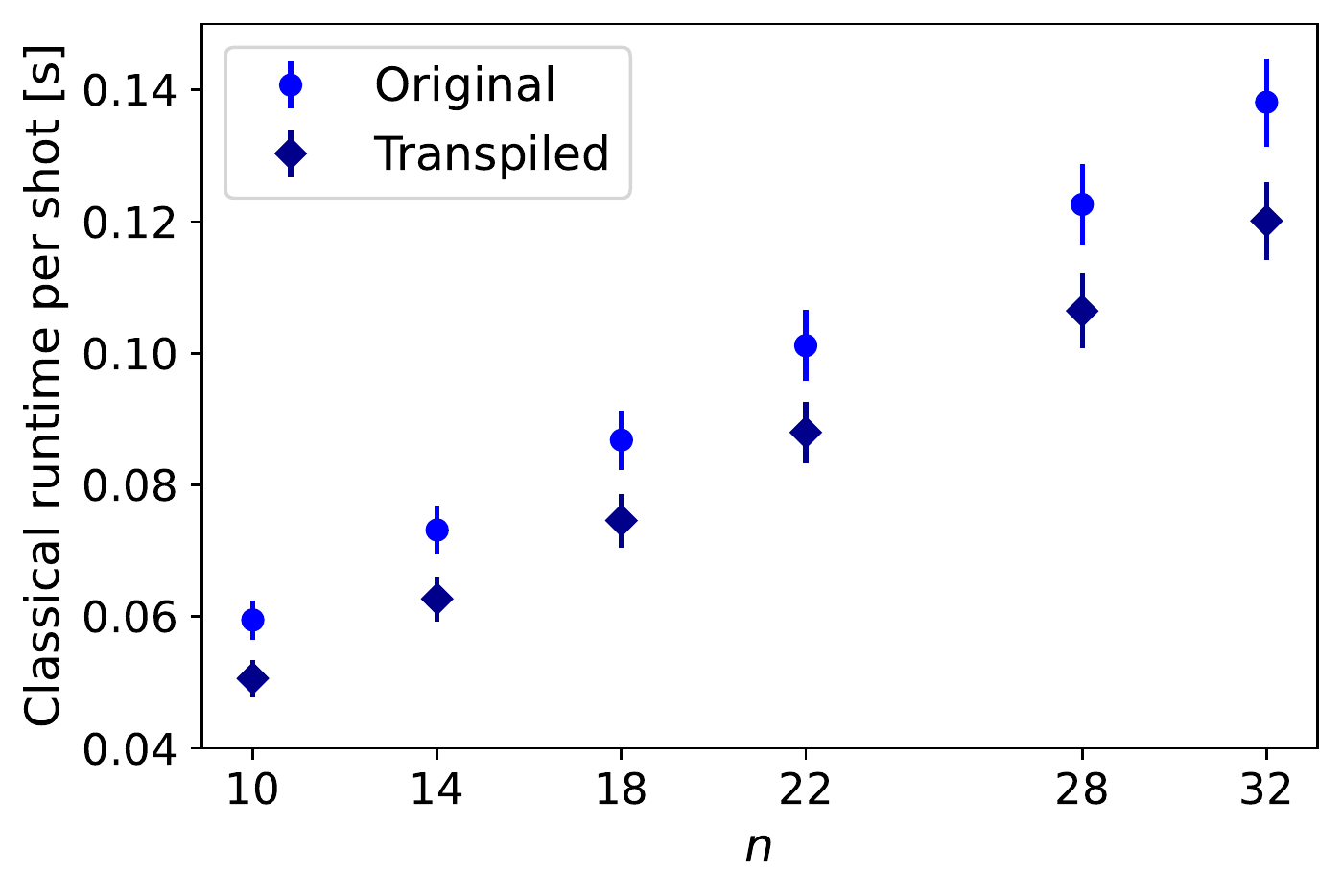}
    \caption{Classical runtime for hidden-shift circuits with $n$ qubits and 14 $T$ gates. Time required for the classical processing that accompanies a single-shot (standard) Pauli-based computation for both the original (light blue circles) and the pre-compiled (dark blue diamonds) hidden-shift circuits featuring 14 $T$ gates, as a function of the number of qubits $n$. The total computational runtime for each PBC shot will be this classical runtime added to the time taken by the QPU.}
    \label{fig: HSCs--classical_runtime}
\end{figure}

\begin{figure}[t]
    \centering
    \includegraphics[scale=0.5]{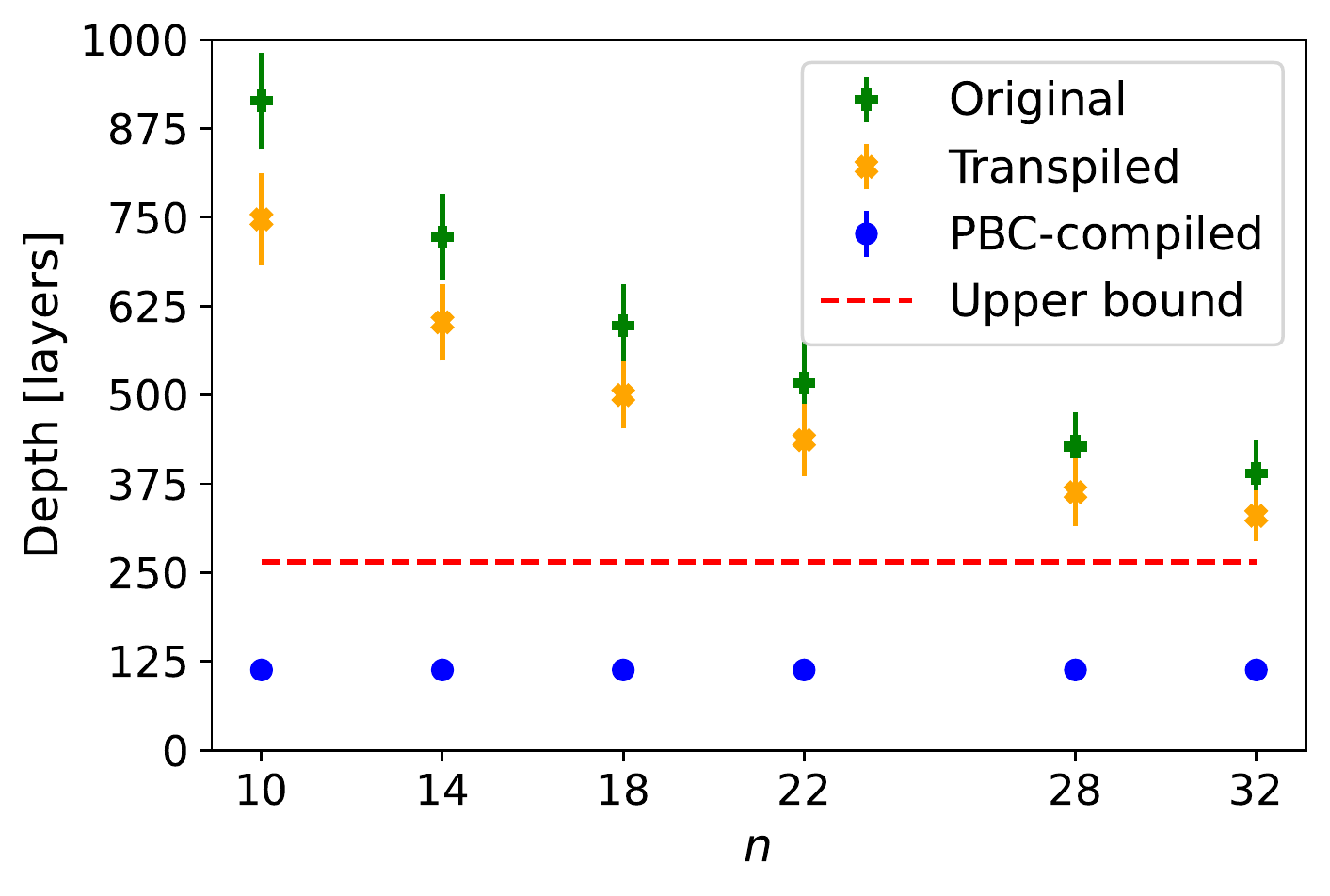}
    \caption{Depth of the original (green), and pre-compiled (orange) hidden-shift circuits with 14 $T$ gates, and of the PBC-compiled circuits (blue), as a function of the number of qubits $n$ of the original circuits. The dashed red line represents the upper bound for the depth of the (standard) PBC ($D^{ub} = t(t+5)-1 = 265$) and, as expected, it is much larger than the actual depths of the compiled circuits.
    }
    \label{fig: HSCs--depths}
\end{figure}

\begin{figure*}[t]
    \centering
    \begin{subfigure}[]{0.48\textwidth}
        \centering
        \includegraphics[scale=0.5]{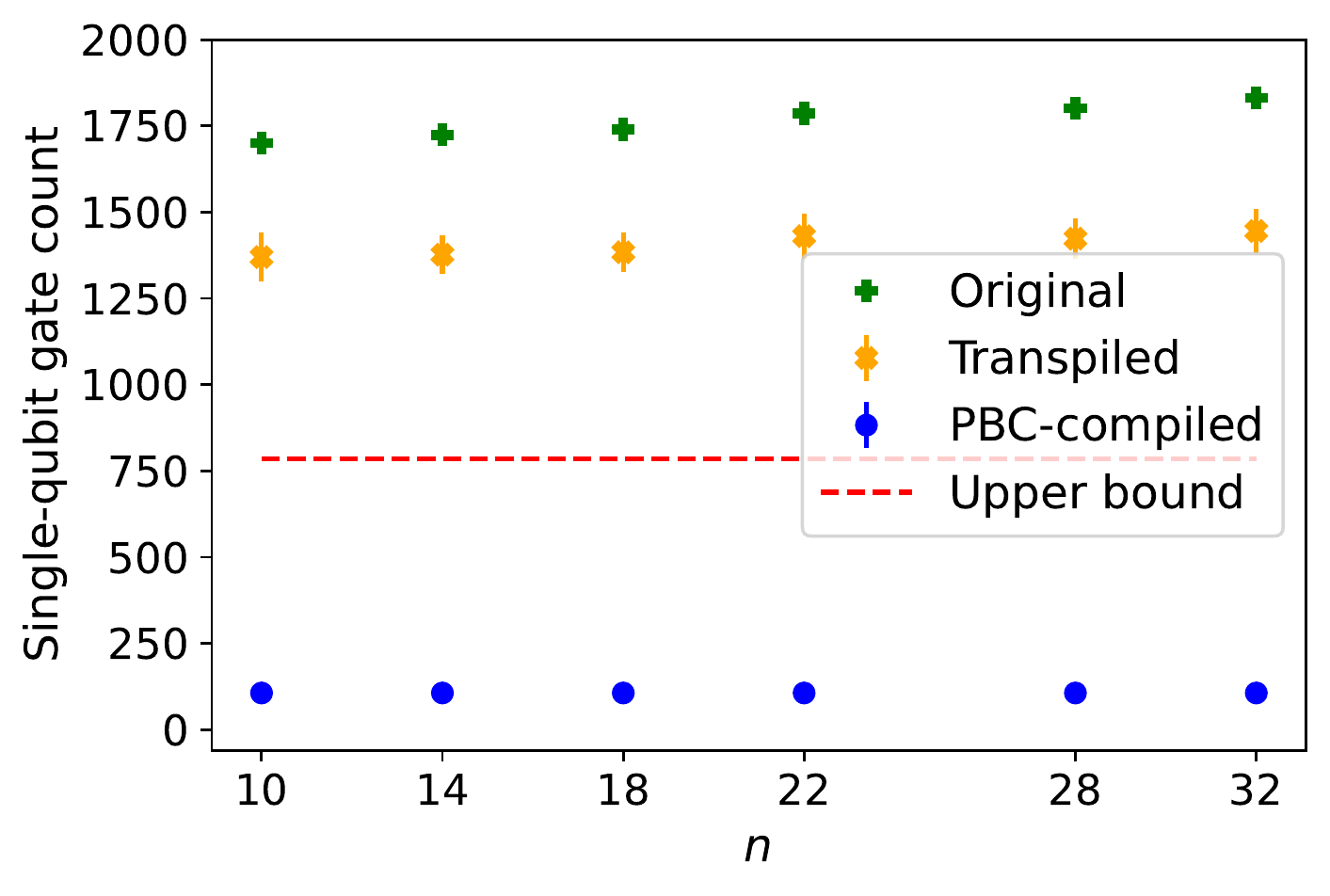}
        \caption{Number of single-qubit gates of the original (green), and pre-compiled (orange) hidden-shift circuits with 14 $T$ gates, and of the PBC-compiled quantum circuits (blue), as a function of the number of qubits $n$ of the original circuits. The dashed red line represents the upper bound for this quantity ($N^{ub}_{HS} = 4t^2 = 784$) and, as expected, it is much larger than the actual final values.}
        \label{subfig: HSCs--HS_count}
    \end{subfigure}
    \hfill
    \begin{subfigure}[]{0.48\textwidth}
        \centering
        \includegraphics[scale=0.5]{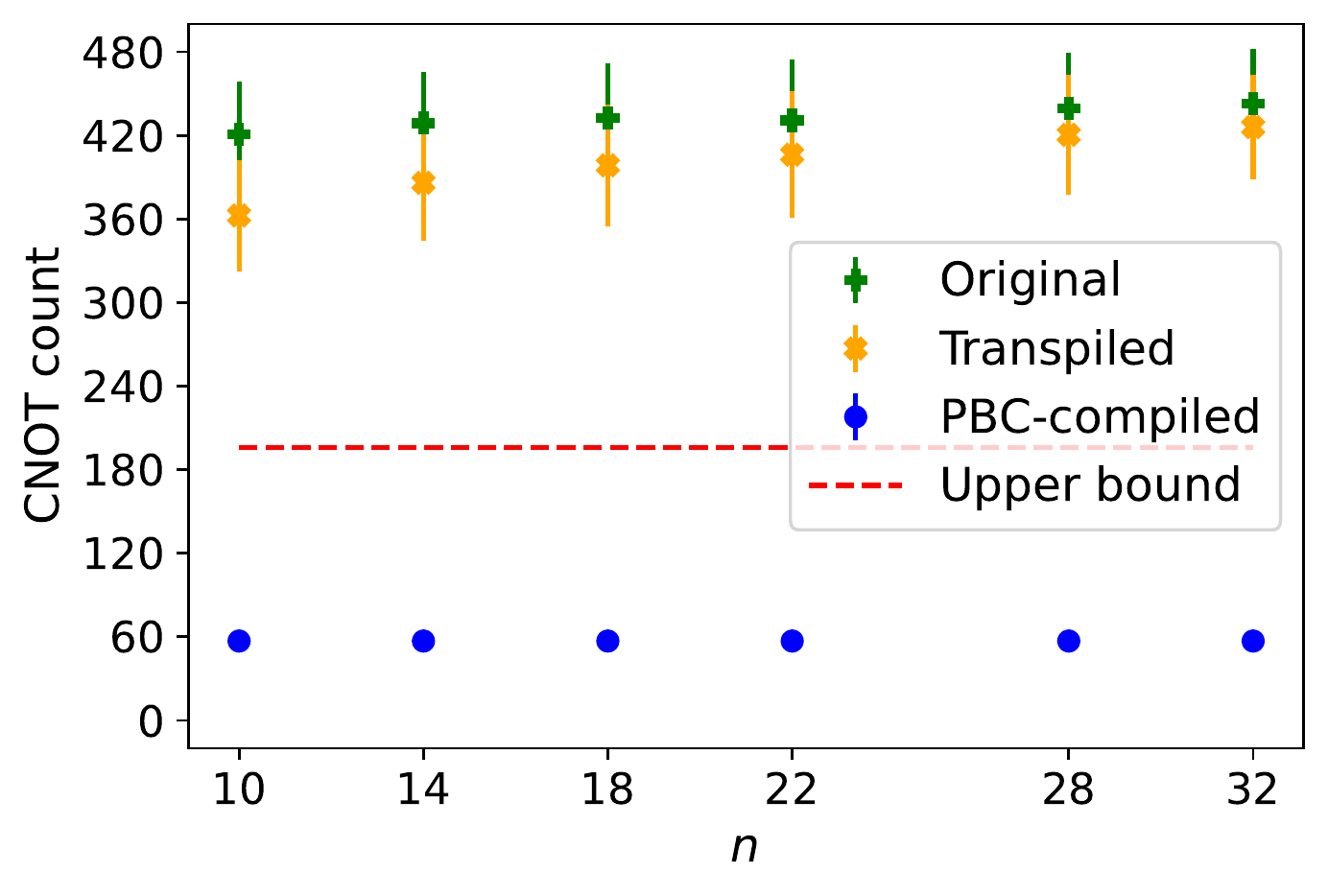}
        \caption{CNOT count of the original (green), and pre-compiled (orange) hidden-shift circuits with 14 $T$ gates, and of the PBC-compiled quantum circuits (blue), as a function of the number of qubits $n$ of the original circuits. The dashed red line represents the upper bound for this quantity ($N^{ub}_{\mathrm{CNOT}} = t^2 = 196$) and, as expected, it is much larger than the actual final values.}
        \label{subfig: HSCs--CNOT_count}
    \end{subfigure}
    \caption{Gate counts for the original (green), and pre-compiled (orange) hidden-shift circuits with 14 $T$ gates, and for the PBC-compiled quantum circuits (blue), as a function of the number of qubits $n$ of the original (input) circuits.}
    \label{fig: Gate counts}
\end{figure*}

The family of HSCs considered in this paper is that of Clifford$+T$ circuits with an even number of qubits, $n$, a hidden string, $\mathbf{s}$, which is returned deterministically, and the following structure:
\begin{equation}
    \mathcal{C} = H^{\otimes n} O_{f^{\prime}} H^{\otimes n} O_f H^{\otimes n}\,,
    \label{eq: HSCs}
\end{equation}
where $O_f$ and $O_{f^{\prime}}$ are the two $n$-qubit oracles of the hidden-shift algorithm given by:
\begin{equation}
    O_f = \left( \prod_{i=1}^{n/2} CZ_{i, i+n/2} \right) \left( O_g \otimes I \right)
    \label{eq: O_f oracle}
\end{equation}
and
\begin{equation}
    O_{f^{\prime}} = \left( \prod_{i=1}^{n/2} CZ_{i, i+n/2} \right) \left( I \otimes O_g \right) Z(\mathbf{s}).
    \label{eq: O_f^prime oracle}
\end{equation}
Here, $Z(\mathbf{s}) = Z^{s_1}\otimes Z^{s_2}\otimes ... \otimes Z^{s_n}$ and $O_g$ is an $n/2$-qubit unitary defined by the number of $CCZ$ gates, $N_{CCZ}$, used to construct it and by the number of $\{ Z,\,CZ\}$ gates, $N_{Z;CZ}$, within each $\{Z,\,CZ\}$-segment placed at the beginning and at the end of $O_g$, and in-between every $CCZ$ gate.\footnote{Note that Equations~\eqref{eq: O_f oracle} and \eqref{eq: O_f^prime oracle}, and the oracle $O_g$ demand that $n$ be an even number.} This structure implies that there are $N_{CCZ}+1$ $\{ Z,\, CZ\}$-segments, each of which is built by drawing $N_{Z;CZ}$ gates uniformly at random from the $\{Z,\,CZ\}$ gate set, and acting with each drawn gate on a randomly chosen qubit or qubits. The three qubits involved in each $CCZ$ are also drawn randomly from the set of $n/2$ qubits acted on by $O_g$.

This family of HSCs is that defined in Ref.~\cite{BG16}, with two major differences. First, after generating the circuits, the $Z$ and $CZ$ gates need to re-written in terms of the (allowed) Clifford generators: $Z_i = S_i^2$ and $CZ_{ij} = H_j CX_{ij} H_j\,.$ Secondly, the $CCZ$ gate can be written in terms of the Toffoli gate, for which we then considered the optimal, non-adaptive decomposition as proposed by Shende and Markov in Ref.~\cite{ShendeMarkov08}. As a consequence, our circuits have a $T$-count which is $t = 14N_{CCZ}$. For this reason, and due to limitations associated with the Schr{\"o}dinger-type simulator, we examined only circuit instances with $t=14$. For more details about this family of circuits, the reader is encouraged to read the Supplementary Material of Ref.~\cite{BG16}.

We used input Clifford$+T$ circuits with different total numbers of qubits: $n=\{10,\, 14,\, 18,\, 22,\, 28,\, 32\},$ and for each $n$ generated a single hidden string $\mathbf{s}$ at random from the $2^n$ possible strings. As explained above, the structure of the generated circuits guarantees that their output is deterministic and corresponds to the generated string: $y_{out} = \mathbf{s}$. Evidently, the same result is output by our code.

For each of the six values of $n$, we generated and simulated $100$ different HSCs. For completeness, we further decided to analyze what would happen if we pre-compiled the input quantum circuits with another tool before using our code. To that end, we used Qiskit's transpiler, although there is nothing special about this particular pre-compilation, and other tools could be used instead before running our code.

We expect that by pre-compiling the quantum circuits, the time of the classical processing is lowered as the number of gates through which the Pauli operators need to be propagated is reduced \textit{a priori}. This is corroborated by the results presented in Figure~\ref{fig: HSCs--classical_runtime}. 

Note that the PBC-compiled (adaptive) quantum circuits require $t+1=15$ qubits, that is, our compilation code yields a reduction of the number of qubits only for the input circuits with $n>16$. In contrast, all other resources, i.e. depth and gate counts, are reduced with respect to those of the original circuits (cf. below).

In Figure~\ref{fig: HSCs--depths}, we see a substantial depth reduction provided by the use of our (adaptive) PBC scheme. Specifically, the compiled quantum circuits have one of five possible depths: 111, 112, 113, 114, or 115. These values are around 2.3 times smaller than the respective upper bound, and between 2.8 and 8.6 times smaller than the depths of the original (input) quantum circuits. 

As for the gate counts, the reductions are also striking (see Figure~\ref{fig: Gate counts}). The total number of single-qubit gates reaches values between 6.5 and 8.5 times smaller than the corresponding upper bound and between 14 and 20 times smaller than those of the original quantum circuits. On the other hand, the total number of controlled-NOT gates in the compiled quantum computations is always either 56, 57 or 58; these values are roughly 3.4 times smaller than the corresponding upper bound, and between 6.6 and 8.6 times lower than those of the original quantum circuits.

All in all, our code achieves some promising reductions of the quantum resources necessary to implement the desired computation, while requiring classical processing that is only of the order of hundreds of milliseconds per shot. These results are encouraging especially in light of the fact that this classical runtime should be significantly improvable. Moreover, it is worth pointing out that even if the input quantum circuits are generated using a larger number of $Z$ and $CZ$ gates, thus having larger depth and gate counts, we still expect the final (adaptive) compiled computation carried out by our code to reach the same numerical results as the ones presented in Figures~\ref{fig: HSCs--depths} and \ref{fig: Gate counts}. As explained at the end of Section~\ref{subsec: Improvement 1: Pauli measurements}, the reason for this is the underlying, strict standard form of the compiled computation, which depends only on the structure of the original circuit and on its $T$-count.

\subsubsection{Random quantum circuits} \label{subsubsec: RQCs}

\begin{figure}[t]
    \centering
    \includegraphics[scale=0.5]{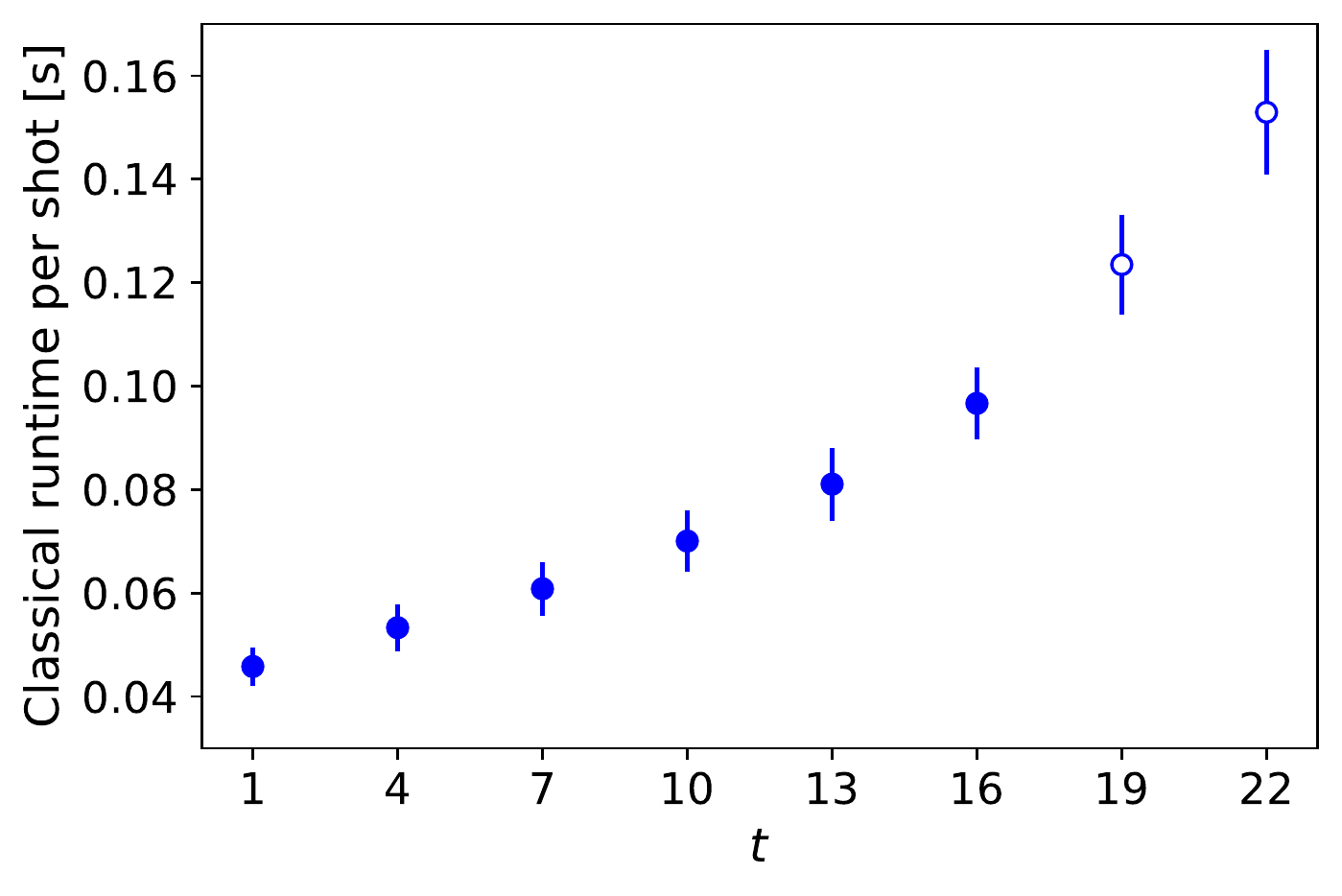}
    \caption{Classical runtime for random quantum circuits with 25 qubits and $t$ $T$ gates. Time required for the classical processing that accompanies a single-shot (standard) Pauli-based computation for random quantum circuits with 25 qubits, as a function of the $T$-count $t$ of the original circuits. The total computational runtime for each PBC shot will be this classical runtime, added to the time taken by the QPU. The last two points are depicted as white-faced circles to highlight the fact that for these two values of $t$ we have simulated only 5 circuits (instead of 100).}
    \label{fig: RQCs--classical_runtime}
\end{figure}

\begin{figure}[t]
    \centering
    \includegraphics[scale=0.5]{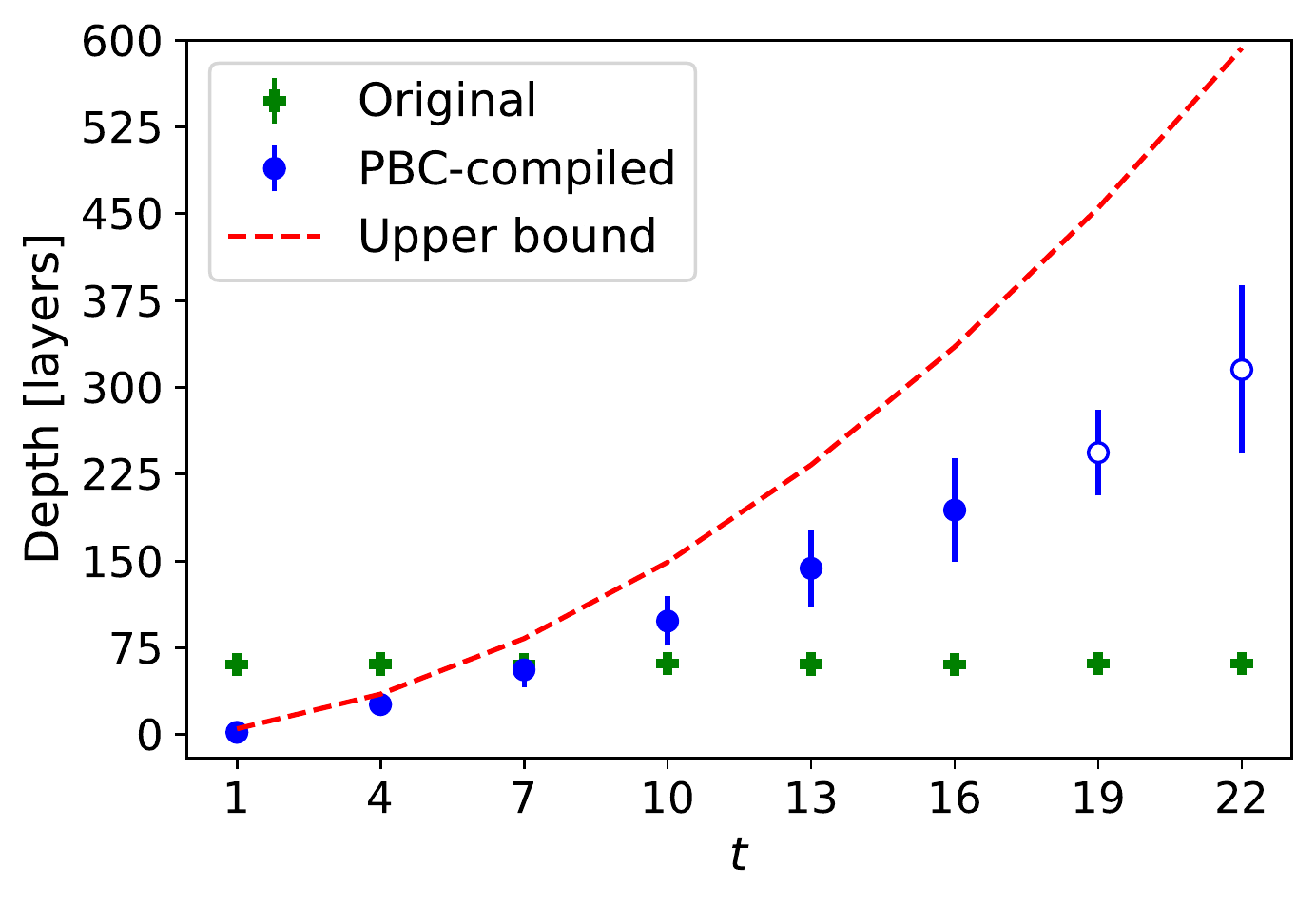}
    \caption{Depth of the input random quantum circuits with 25 qubits (green) and of the compiled quantum circuits (blue), as a function of the $T$-count, $t$, of the original circuits. The dashed red line represents the upper bound for the depth of the PBC ($D^{ub} = t(t+5)-1$). The last two points are depicted as white-faced circles to highlight the fact that for these two values of $t$ we have simulated only 5 circuits (instead of 100).}
    \label{fig: RQCs--depths}
\end{figure}

\begin{figure*}[t]
    \centering
    \begin{subfigure}[t]{0.48\textwidth}
        \centering
        \includegraphics[scale=0.5]{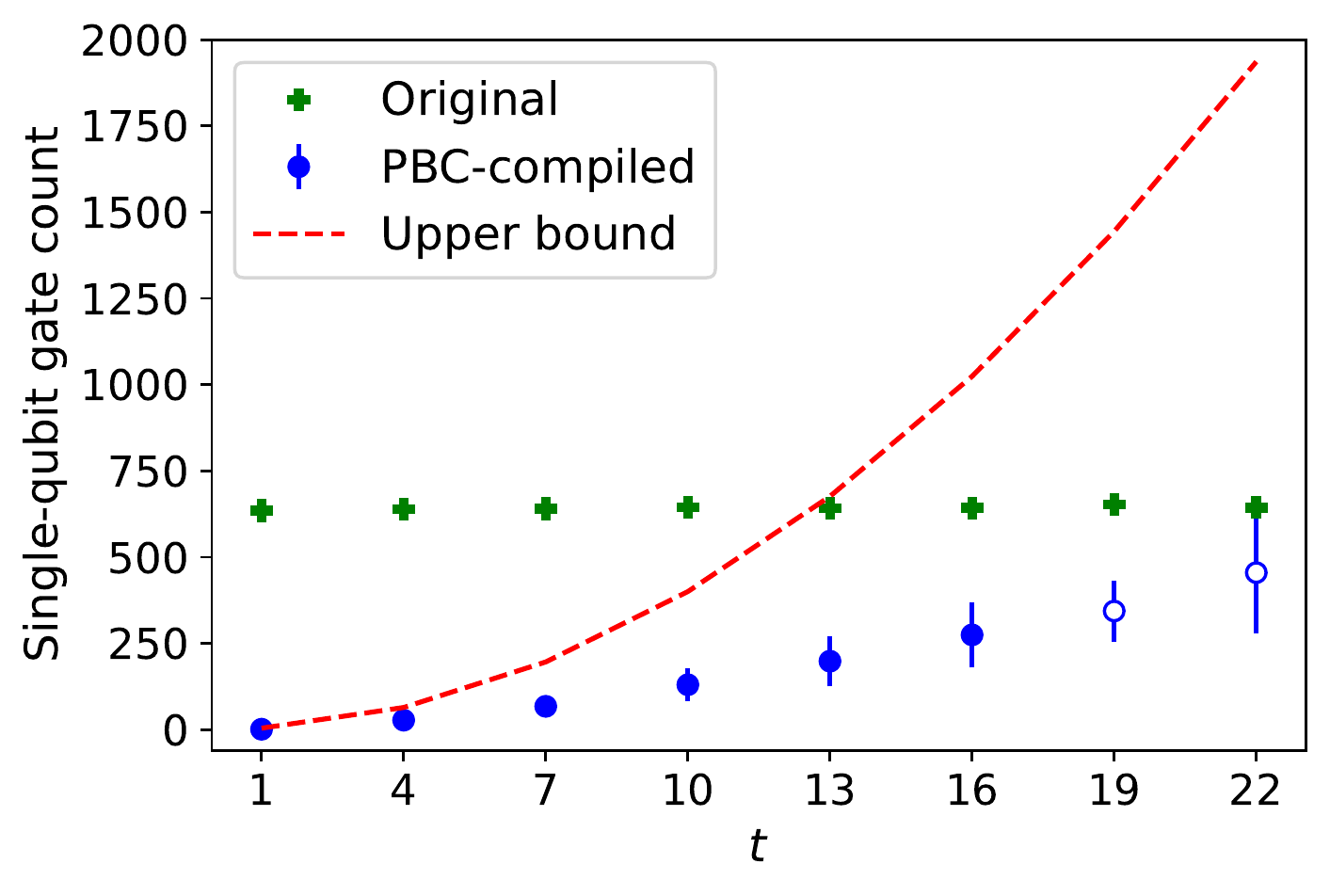}
        \caption{Number of single-qubit gates of the input 25-qubit random quantum circuits (green) and of the (respective) compiled quantum circuits (blue), as a function of the number of $T$ gates, $t$, of the original circuits. The dashed red line represents the upper bound for this quantity ($N^{ub}_{HS} = 4t^2$). The last two points are depicted as white-faced circles to highlight the fact that for these two values of $t$ we have simulated only 5 circuits (instead of 100).}
        \label{subfig: RQCs--HS_count}
    \end{subfigure}
    \hfill
    \begin{subfigure}[t]{0.48\textwidth}
        \centering
        \includegraphics[scale=0.5]{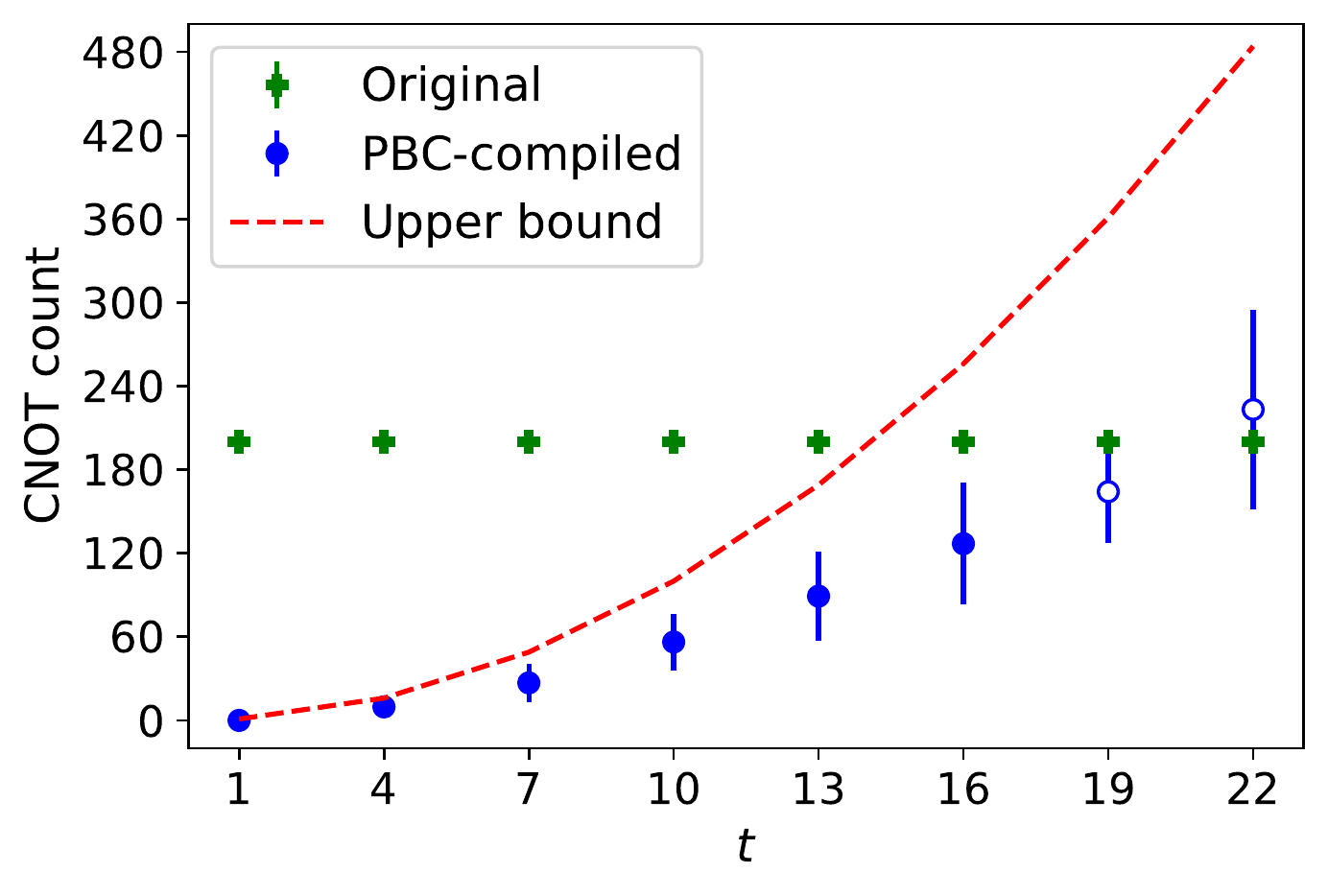}
        \caption{CNOT count of the input 25-qubit random quantum circuits (green) and of the (respective) compiled quantum circuits (blue), as a function of the number of $T$ gates, $t$, of the original circuits. The dashed red line represents the upper bound for this quantity ($N^{ub}_{\mathrm{CNOT}} = t^2$). The last two points are depicted as white-faced circles to highlight the fact that for these two values of $t$ we have simulated only 5 circuits (instead of 100).}
        \label{subfig: RQCs--CNOT_count}
    \end{subfigure}
    \caption{Gate counts of the original 25-qubit random quantum circuits (green) and of the PBC-compiled quantum circuits (blue), as a function of the $T$-count, $t$, of the original circuits.}
    \label{fig: RQCs -- Gate counts}
\end{figure*}

Having verified that our code works flawlessly with the HSCs, outputting the correct deterministic result, we now turn our attention to the case of random input Clifford$+T$ quantum circuits. To complement the data obtained with the HSCs, we generated random circuits with a fixed number of qubits $n=25$, but varying $t=\{1,\, 4,\, 7,\, 10,\, 13,\, 16,\, 19,\, 22\}.$ This generation was strongly inspired by the random quantum circuits (RQCs) proposed in Ref.~\cite{Google18}:

\begin{enumerate}
    \item The $n=25$ qubits are initialized in the state $\left| 0 \right>^{\otimes n}$.
    \item The first layer of the circuit consists of Hadamard gates acting on all qubits: $H^{\otimes n}$.
    \item The next $N_{cycles}$ layers are generated by picking up $1$ out of the $8$ possible entangling configurations presented in Figure~2a of Ref.~\cite{Google18}. This is motivated by practical constraints in qubit connectivity, assuming a device that has a rectangular-grid configuration so that the $n$ qubits are arranged in an $(n_c \times n_r)$-grid. This means that $n=n_c\cdot n_r\,,$ where $n_c$ and $n_r$ denote, respectively, the number of columns and rows in the grid. Our circuits assume a $(5\times 5)$-grid structure and were generated with $N_{cycles}=40.$ Note that the entangling gate used is the controlled-$Z$ gate. 
    \item For each layer, each qubit that is not acted on by a $CZ$ gate will be acted on by a single-qubit gate drawn from the $\{ S,\, H,\, T \}$ gate set. The drawing of such gates should follow the ensuing rules:
    \begin{enumerate}
        \item the probability $p(T)$ of drawing $T$ is adjusted so that, in the end, the total number of $T$ gates is (close or equal to) the desired one;
        \item the probability of drawing either one of the other two gates is the same and given by $p(S)=p(H)= (1-p(T))/2\,.$
    \end{enumerate}
    A post-selection criterion is placed on the generated samples to ensure that they have the desired number of $T$ gates.
    \item In the end, all qubits are measured in the computational basis so that $m=n=25.$
\end{enumerate}

Note that unlike the set-up in Ref.~\cite{Google18}, here we do not demand that the first single-qubit gate acting on each qubit (after the initial $H^{\otimes n}$ layer) be a $T$ gate. This is important because it allows more flexibility in the number of $T$ gates in the circuit. Additionally, and also unlike Ref.~\cite{Google18}, our set-up might allow the commutation of some gates and the simplification of the quantum circuits.
For this reason, two more steps were taken before feeding these circuits as input to our code:
\begin{enumerate}
    \item The entangling $CZ$ gates were transformed into CNOTs.
    \item We made a pre-compilation of these circuits, using gate commutativity and circuit identities to remove disposable gate sequences. After this, the total $T$-count of the circuit must remain the desired one, or else that circuit is discarded.
\end{enumerate}

Here, the values of $n$ and $t$ imply that all RQCs are compiled to instances with lower qubit counts. Just like for the HSCs, we simulated 100 different RQCs for $t$ from 1 up to 16. However, for the last two $T$-count values, time constraints (related to the Statevector simulator) led us to simulate only 5 different random circuits. For this reason, the corresponding data points in the figures are distinguished from the others by being represented by empty circles (instead of filled ones).

Figure~\ref{fig: RQCs--classical_runtime} depicts the classical time associated with the compilation procedure as a function of the number of non-Clifford gates in the original circuits. Just as before, this time is around hundreds of milliseconds per shot, but we expect this to be significantly improvable.

We can see the depth associated with these compiled circuits in Figure~\ref{fig: RQCs--depths}. In this case, and contrary to what happened with the HSCs, we see that for $T$-counts larger than $t=7$ the depth of the PBC-compiled computations is actually larger than that of the original circuits. However, it is important to notice that if we use a larger number of layers, $N_{cycles}$, we will increase the depth of the original computation without having any effect on the depth of the final (PBC-compiled) circuits.

We also see, from Figure~\ref{fig: RQCs -- Gate counts}, that despite their larger depth for $t>7\,,$ the total gate counts of the PBC-compiled circuits are still smaller than those of the original Clifford$+T$ quantum circuits. (The exception to this is the CNOT count for $t=22$.) This shows that in terms of the quantum gates required for the actual quantum computation, our approach is still favorable, although it might be deeper due to the particular form of the implementation scheme we are using. In fact, these results suggest that these RQCs are instances that might benefit significantly from the use of our second scheme with the logarithmic-depth CNOT cascade (Section~\ref{subsubsec: 2nd proposed scheme -- t*logt depth}) and also potentially from the space/time trade-off provided by the scheme proposed in Section~\ref{subsubsec: 3rd proposed scheme -- linear depth}.

Finally, we should once more reinforce the idea that if the original quantum circuits are made to be larger (for instance due to an increase of the grid size and/or of the number of cycles), the size of the compiled circuits will remain essentially unchanged as long as the value of $t$ is maintained.

\subsubsection{Dummy simulation and larger circuits} \label{subsubsec: Dummies}

\begin{figure*}
    \centering
    \begin{subfigure}[t]{0.48\textwidth}
        \centering
        \includegraphics[scale=0.5]{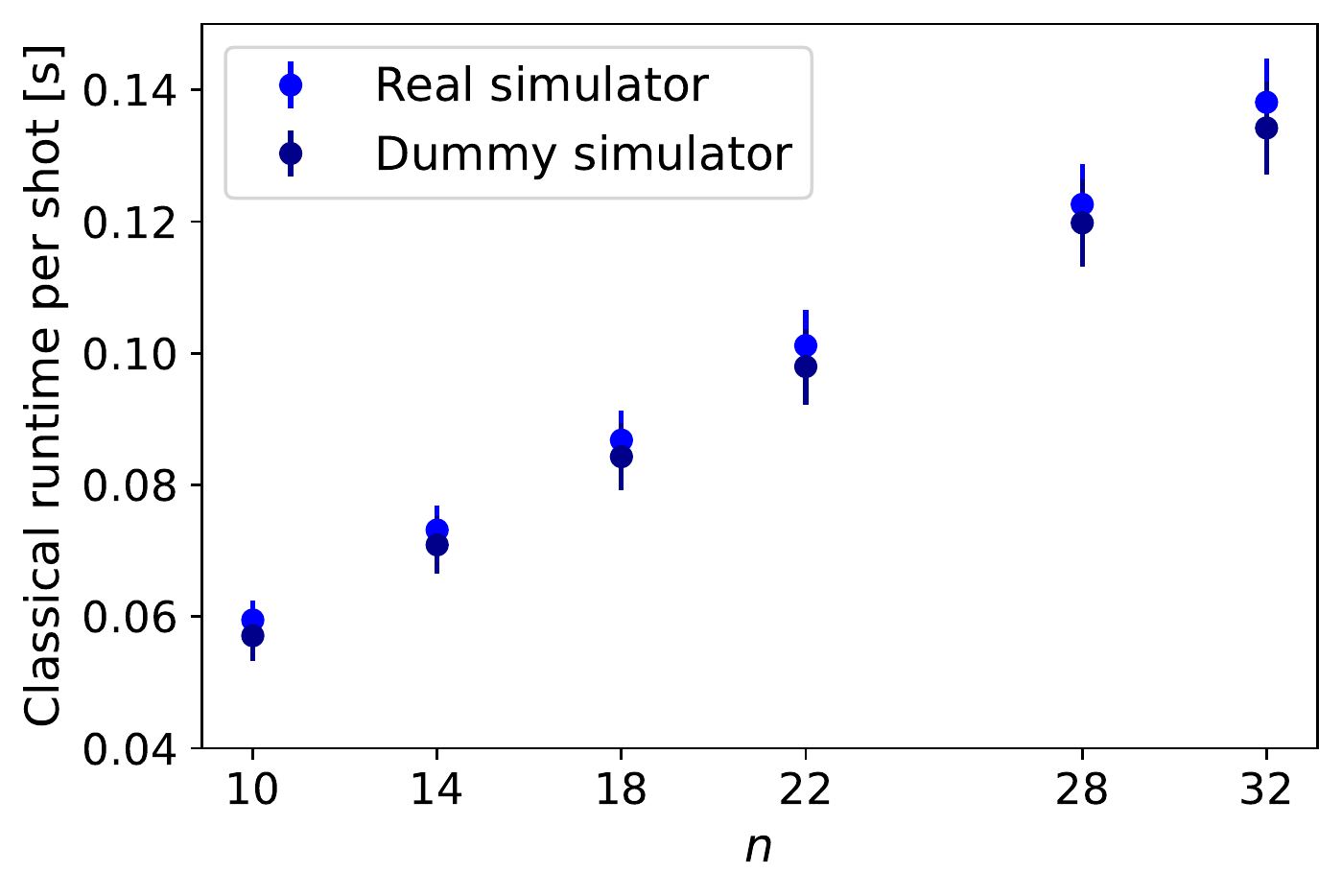}
        \caption{Classical runtime for hidden-shift circuits with $n$ qubits and 14 $T$ gates. Time required for the classical processing that accompanies a single-shot (standard) Pauli-based computation for the original hidden-shift circuits featuring 14 $T$ gates, as a function of the number of qubits $n$, when using the (real) Schr{\"o}dinger-type simulator (light blue) and the dummy simulator (dark blue).}
        \label{subfig: Dummy_v_real--cl_time--HSCs}
    \end{subfigure}
    \hfill
    \begin{subfigure}[t]{0.48\textwidth}
        \centering
        \includegraphics[scale=0.5]{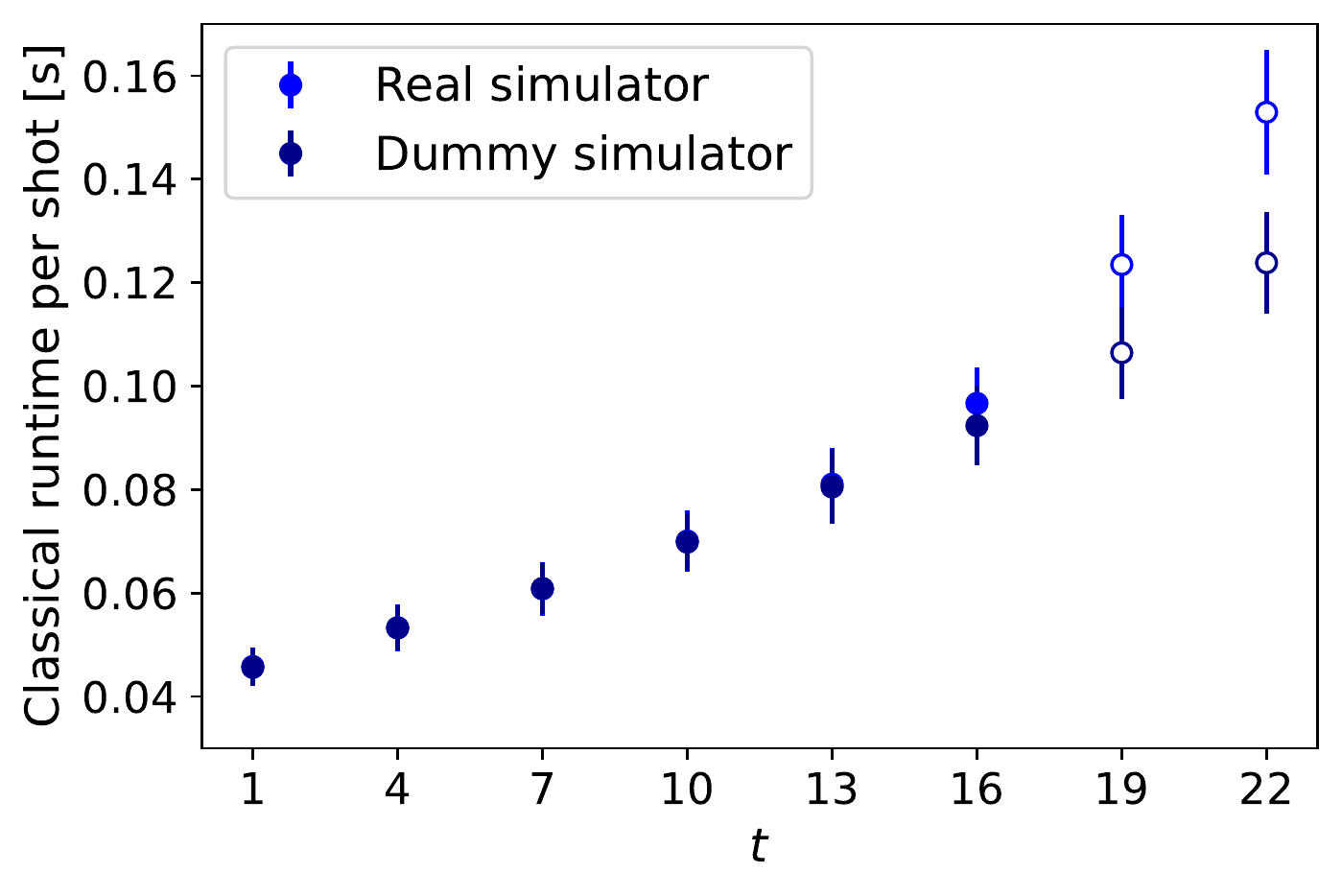}
        \caption{Classical runtime for random quantum circuits with 25 qubits and $t$ $T$ gates. Time required for the classical processing that accompanies a single-shot (standard) Pauli-based computation for random quantum circuits with 25 qubits, as a function of the $T$-count, $t$, of the original circuits, when using the (real) Schr{\"o}dinger-type simulator (light blue) and the dummy simulator (dark blue). The last two data-points are depicted as white-faced circles to highlight the fact that for these two values of $t$ we have simulated only 5 circuits (instead of 100).}
        \label{subfig: Dummy_v_real--cl_time--RQCs}
    \end{subfigure}
    \caption{Time of the classical processing for (a) a set of hidden-shift circuits and (b) a set of random quantum circuits, when using the (real) Schr{\"o}dinger-type simulator (light blue) and the dummy simulator (dark blue).}
    \label{fig: Dummy_v_real--cl_runtime}
\end{figure*}

In order to verify what happens for larger quantum circuits, we included in our code the possibility of running a \textit{dummy} simulation of the actual quantum measurements of the standard PBC. This dummy simulation works by simply making a coin toss to ascertain the outcome of each Pauli observable that should be measured in the QPU. It is clear that, in doing so, we are not making a correct weak simulation of the original circuit, i.e. we are not sampling from the correct output distribution. This is because we will explore paths down the PBC tree that would not occur in the real simulation, or that would occur with different probability.

Nevertheless, using this dummy simulator allows us to use our code to compile quantum circuits with larger $T$-counts than those allowed by the use of the real Schr{\"o}dinger-type simulator and, thus, get a prediction for the features of the PBC-compiled circuits (depth and gate counts), as well as for the time of the classical processing for those instances. Despite the fact that the PBC tree will not be explored correctly, yielding a wrong weak simulation of the original circuit, we have verified that this dummy simulator reliably estimates the final features of the compiled quantum circuits. We did this by simulating the same hidden-shift and random quantum circuits as compiled with the real simulator in Sections~\ref{subsubsec: HSA} and \ref{subsubsec: RQCs}. 

In the case of the HSCs, we clearly see the effect of sampling from the incorrect probability distribution, as the output is no longer deterministic. Nevertheless, the results for the depth and gate counts were (essentially) the same as obtained with the real Schr{\"o}dinger-type simulation. The same holds for the results associated with the RQCs.

On the other hand, it was interesting (and somewhat surprising) to observe that the estimates of the classical runtime are different when we use the real, Schr{\"o}dinger-type simulator and the dummy simulator. This can be seen in Figure~\ref{fig: Dummy_v_real--cl_runtime}. For the HSCs, the differences are small (between 2 and 4ms) and negligible, since they are lower than the width of the confidence intervals; they also seem to be approximately constant (or else varying quite slowly) with  $n$ (cf. Figure~\ref{subfig: Dummy_v_real--cl_time--HSCs}). In contrast, for the RQCs, as the $T$-count increases, the discrepancy between the classical runtime per shot estimated using the Schr{\"o}dinger-type or the dummy simulator also increases (cf. Figure~\ref{subfig: Dummy_v_real--cl_time--RQCs}).

From our point of view, there might be two factors at play, contributing to these observed differences. First, they may be precisely due to the fact that different PBC paths are run through with incorrect probability when using the dummy simulator, requiring different computational efforts. However, we believe that such should be a very small contribution, of only a couple of milliseconds, as observed for instance for the HSCs. The second, and probably larger factor, comes from the overwork incurred on the classical machine due to the highly demanding Schr{\"o}dinger-type circuit simulation. One strong argument to support this claim is the fact that, contrary to what happens with $n$, the discrepancies increase visibly with $t$, which is the variable that determines the hardness of the simulation for the real, Schr{\"o}dinger-type simulator. For this reason, we expect that the time of the classical processing is better, even if not perfectly, estimated by the use of the dummy simulator. 

Since the dummy simulator can be used to reliably estimate the properties of these PBC-compiled quantum circuits, we decided to employ it in our code to study the results of the compilation procedure when applied to input quantum circuit instances with larger numbers of $T$ gates.

We started by using this set-up to compile a set of 50 HSCs with 42 qubits and 42 $T$ gates, generated as explained in Section~\ref{subsubsec: HSA}. The classical processing time per shot required to compile one of these circuits can be found within the $95\%$ confidence interval $\tau_{cl.} = ( 0.58\pm 0.02) \mathrm{s}$. Clearly, this value is too demanding for current NISQ devices; however, as has been mentioned several times, this classical processing time should be significantly improvable.

Regarding the properties of the original and compiled quantum circuits, we start from instances with depth $d_{orig.}=(568\pm 64)$ and these are compiled to circuits with depth $d_{comp.} = (357\pm 25)$; this corresponds to reductions in the depth between 24\% and 48\%. For the gate counts, we start from circuits with $3\,572$ single-qubit gates, and end up with compiled circuits with $N_{HS;comp.} = 350\pm 47$, which corresponds to a reduction between 89\% and 92\%. Finally, the CNOT gate count changes from $N_{\mathrm{CNOT};orig.} = 881\pm 54$ to $N_{\mathrm{CNOT};orig.} = 187\pm 26,$ amounting to a reduction between 74\% and 83\%.

\begin{figure}[t]
    \centering
    \includegraphics[scale=0.52]{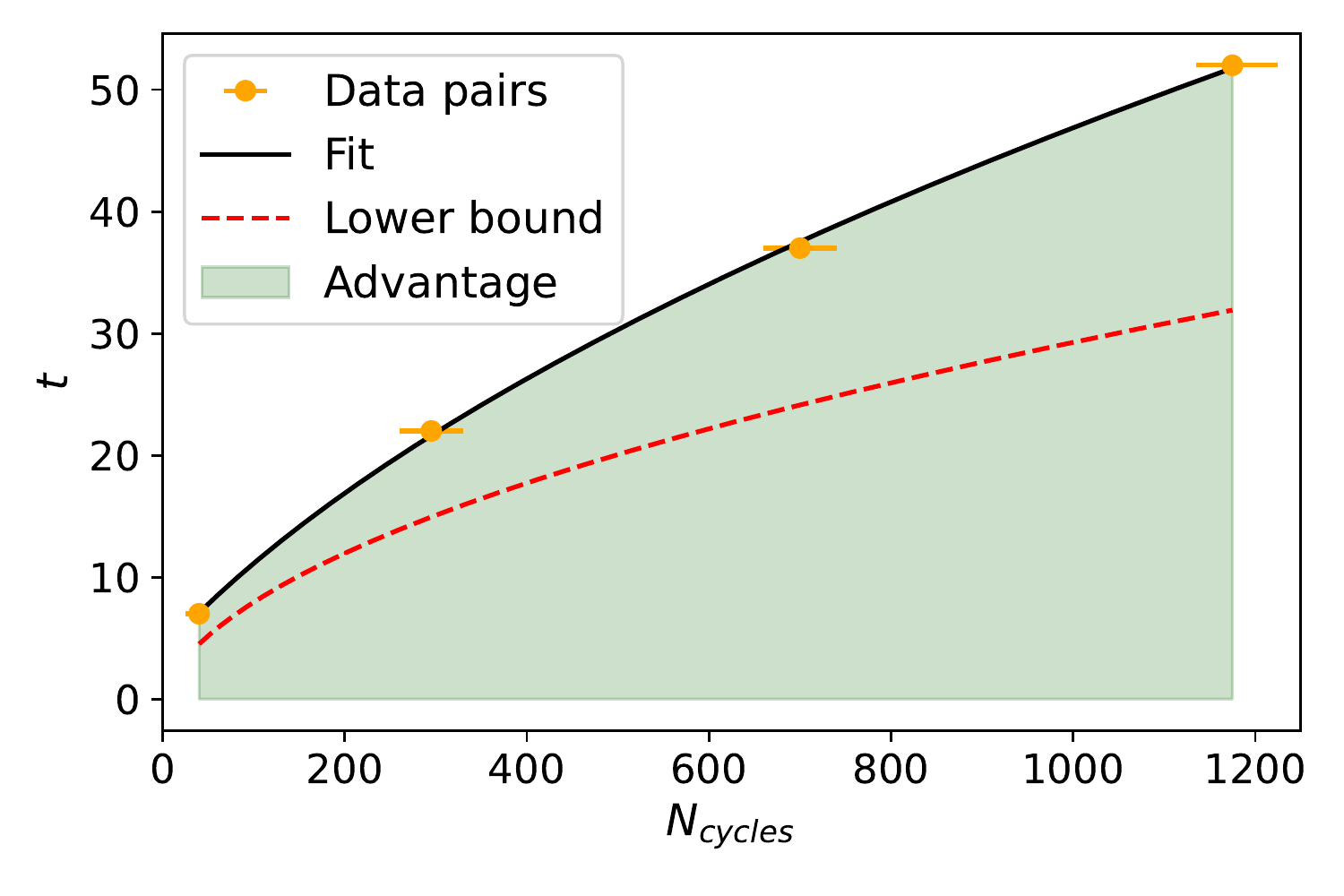}
    \caption{Region of advantage of PBC compilation for random quantum circuits on $25$-qubits, with respect to depth. The orange data points represent the $(N_{cycles}, \, t)$-pairs for which the respective random quantum circuits and their PBC compilation have the same depth. These points result from compiling circuits with $(N_{cycles},\,t)=\{ (22,\, 7),\, (295,\, 22),\, (700,\, 37),\, (1175,\, 52)\}$ using the dummy simulation option in our code. The black line  is obtained by fitting a square-root function to the four data points. This boundary line is consistent with the lower bound (dashed red curve) determined in the main text. The green region corresponds to random-circuit parameters for which PBC compilation decreases the depth, whereas circuits in the white region have their depth increased by PBC compilation.
    }
    \label{fig: Boundary}
\end{figure}

\begin{figure*}[t]
    \centering
    \begin{subfigure}[]{0.48\textwidth}
        \centering
        \includegraphics[scale=0.49]{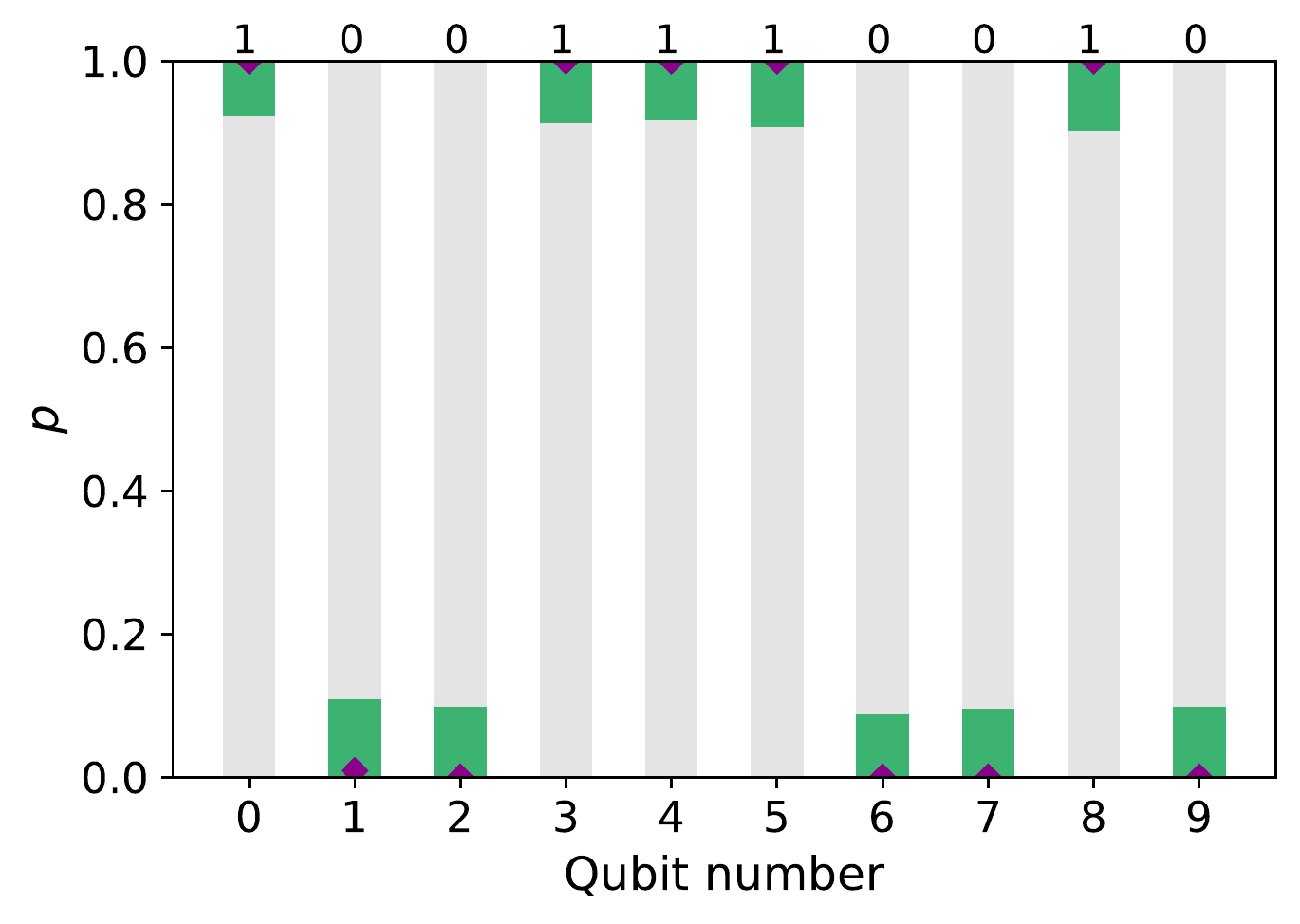}
        \caption[]{1 virtual qubit}
        \label{subfig: Hybrid-HS0-1vq}
    \end{subfigure}
    \hfill
    \begin{subfigure}[]{0.48\textwidth}
        \centering
        \includegraphics[scale=0.49]{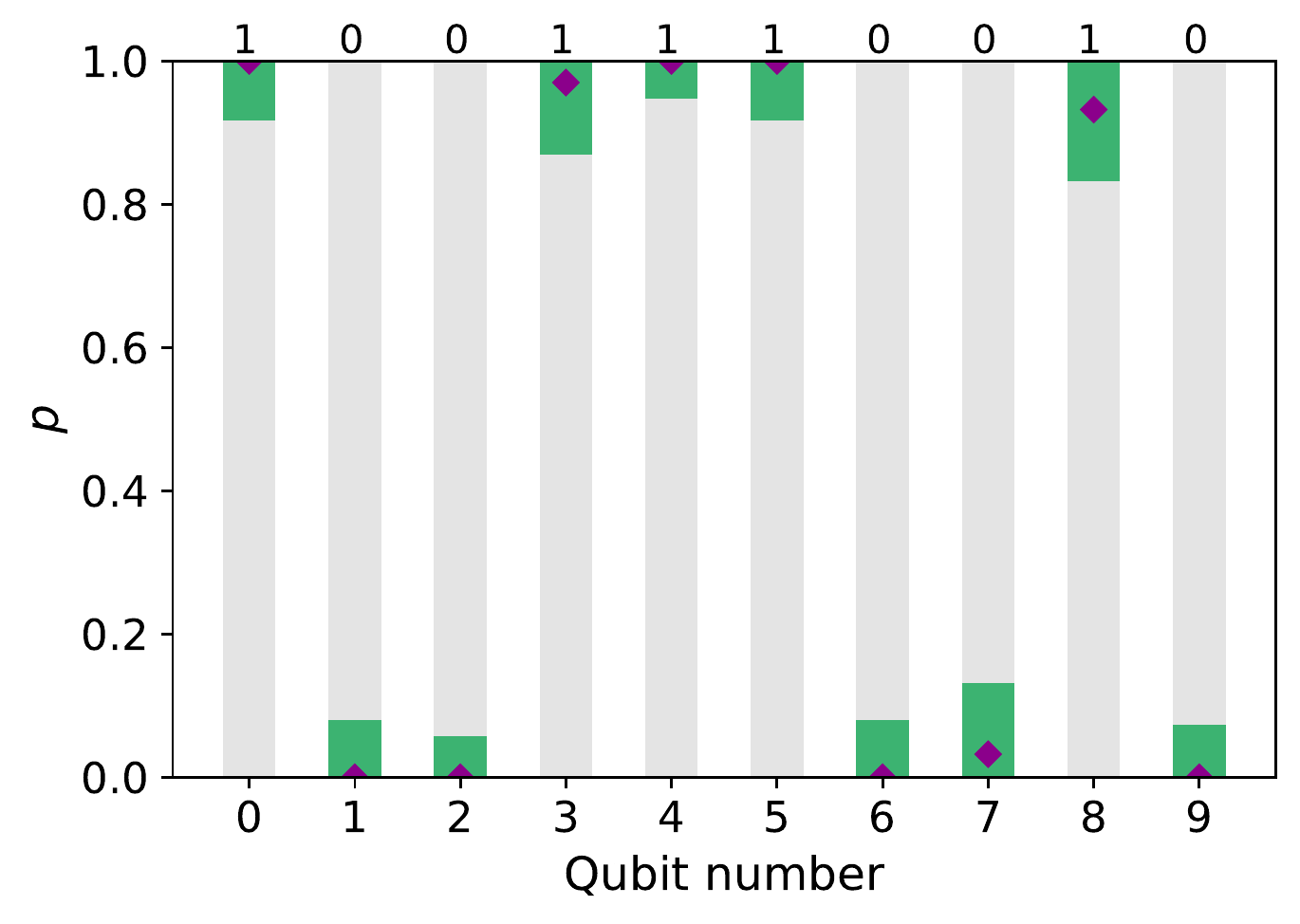}
        \caption[]{2 virtual qubits}
        \label{subfig: Hybrid-HS0-2vq}
    \end{subfigure}
    \newline
    \begin{subfigure}[]{0.48\textwidth}
        \centering
        \includegraphics[scale=0.49]{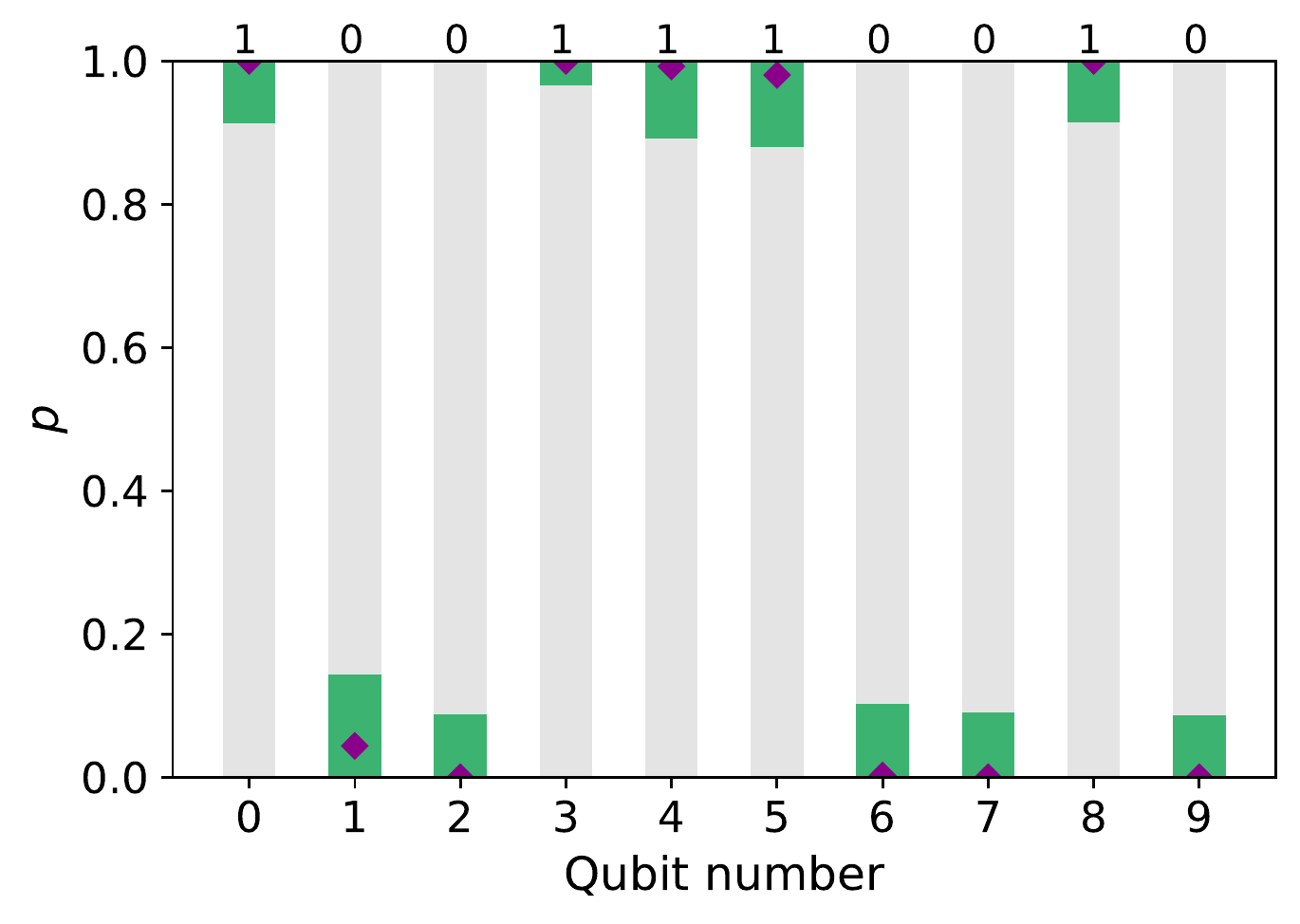}
        \caption[]{3 virtual qubits}
        \label{subfig: Hybrid-HS0-3vq}
    \end{subfigure}
    \hfill
    \begin{subfigure}[]{0.48\textwidth}
        \centering
        \includegraphics[scale=0.49]{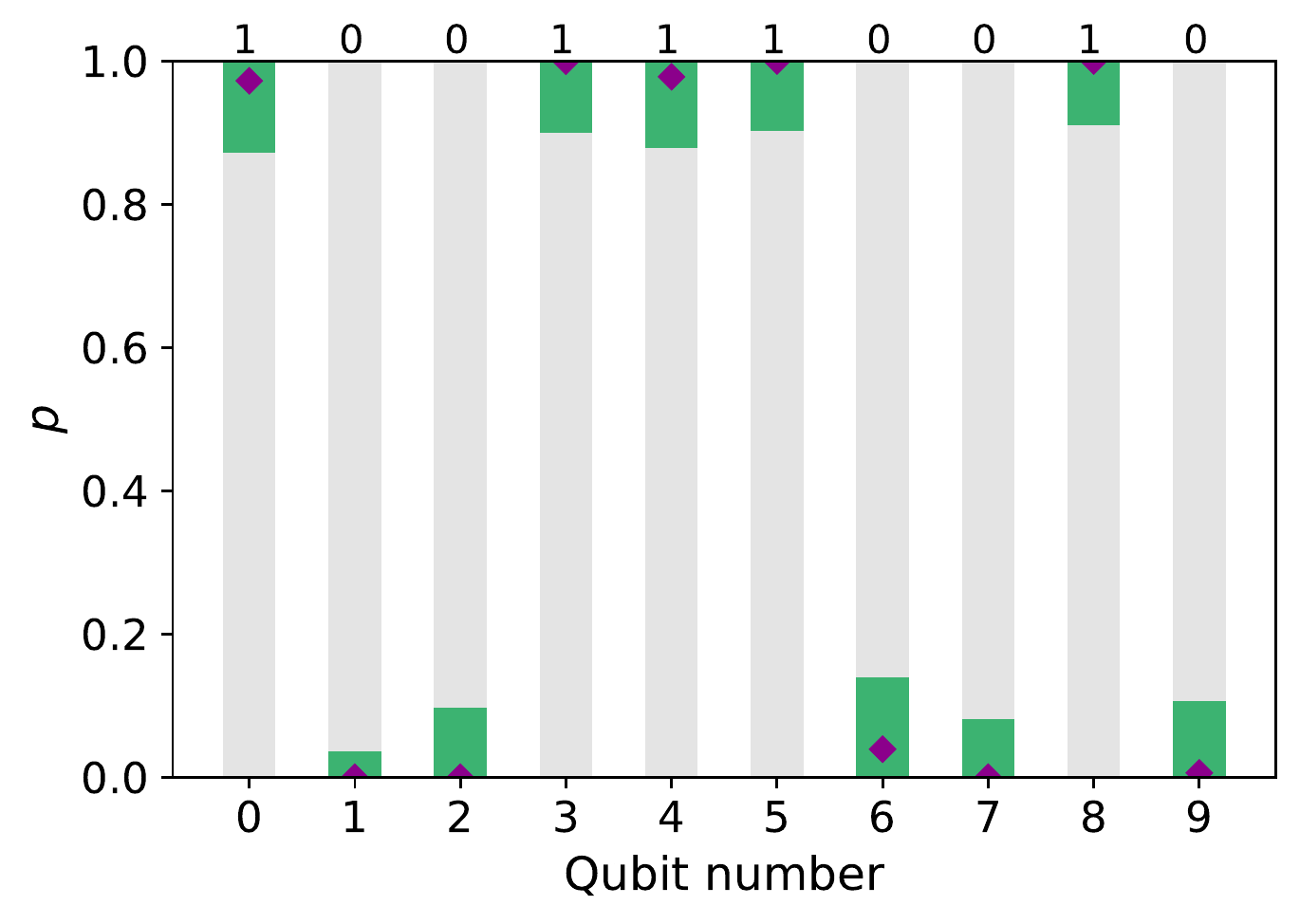}
        \caption[]{4 virtual qubits}
        \label{subfig: Hybrid-HS0-4vq}
    \end{subfigure}
    \caption{Probability of getting the outcome 1 for each of the 10 qubits of a (randomly chosen) hidden-shift circuit with 14 $T$ gates, for 4 different numbers of virtual qubits. The precision was set at $\epsilon=0.1$ so that the total number of shots (per qubit) is $\mathcal{N}= 530,\,\, 1\,060,\,\, 2\,120,$ and $4\,239$, respectively for 1, 2, 3 and 4 virtual qubits (cf. Table~\ref{tab: Hybrid_PBC-nr_samples}). The magenta diamonds represent the expectation value of the random variable $\eta$ and the green bars correspond to 99\% confidence intervals following Hoeffding's inequality. At the top of each bar, we see the correct deterministic output for each qubit. We note that all estimates of $p$ are consistent with the correct output.}
    \label{fig: Hybrid--HSC}
\end{figure*}
These results show that our PBC compilation is very successful in reducing the quantum resources required to carry out the computation encoded by the hidden-shift circuits of the considered family, both for smaller and larger instances.

Next, we decided to consider again the case of random quantum circuits with $n=25$. As explained before, input circuits with a larger number of qubits will have the same depth (as long as $N_{cycles}$ is unchanged) and larger gate counts; on the other hand, the compiled circuits will have essentially the same properties, as the PBC structure is insensitive to changes in the total number of qubits in the grid. For $n=25$ and $N_{cycles} = 40$, we have seen that the compiled circuits start being deeper than the original ones for $t\gtrsim 7$. Using the dummy simulator, we compiled circuit instances with larger $t$ and $N_{cycles}$ to set up an approximate boundary line which determines for each $N_{cycles}$ the value of $t$ below which the PBC compilation leads to circuits that are shallower than the original one.

We can easily derive a lower bound for this boundary line. To that end, recall that the upper bound on the depth of the PBC-compiled circuits is such that $d_{PBC}\leq t^2+5t-1$ (cf. Section~\ref{sec: Main contributions}). Next, we note that, when written in terms of the Clifford generators, the depth of the original Clifford$+T$ RQCs necessarily obeys the condition $d_{orig.}\geq N_{cycles} + 2.$ Combining these two conditions, we get a lower bound for the boundary line such that:
\begin{equation}
    t \geq -\frac{5}{2} + \frac{1}{2} \sqrt{4N_{cycles} + 37}\,.
    \label{eq: lower bound - boundary line}
\end{equation}
Thus, we have $t\sim \Omega(\sqrt{N_{cycles}}).$ More than that, we still expect the boundary line to be such that $t\sim \sqrt{N_{cycles}}$, albeit with coefficients that bring it (necessarily) above the lower bound.

These predictions seem to be corroborated by Figure~\ref{fig: Boundary}. To obtain this Figure, we compiled four sets of 100 quantum circuits with $(N_{cycles},\,t)=\{ (22,\, 7),\, (295,\, 22),\, (700,\, 37),\, (1175,\, 52)\}$. For each of these pairs, the average depth of the PBC-compiled circuits matches that of the original circuits. This gave us the orange data points depicted in Figure~\ref{fig: Boundary}. From the previous sub-subsection, we understand that the depth of the PBC-compiled circuit is better defined by a confidence interval than a single value. The lower and upper bounds of that interval allow us to find lower and upper bounds for the value of $N_{cycles}$ of the original circuits which are still compatible with the confidence interval of the depth of PBC-compiled circuits. This gave us the uncertainty bars depicted in orange in Figure~\ref{fig: Boundary}.

The black line depicted in this same Figure was constructed by fitting a square-root curve to the four data points. This line corresponds to the set of $(N_{cycles}, \, t)$-pairs for which PBC compilation does not change the depth with respect to that of the original circuits. As expected, the quality of the fit is good, meeting our expectation that $t\sim \sqrt{N_{cycles}}$; additionally, the curve respects the derived lower bound (dashed red line).

The green region highlights the set of $(N_{cycles}, \, t)$-pairs that define random circuit instances whose PBC-compiled quantum circuits will be shallower than the original. Note that these results should be independent of the number of qubits in the original RQCs.

To conclude, we would like to note that the classical runtime per shot for the larger RQCs was $\tau_{cl.}=(2.5\pm 0.2)\mathrm{s}$. By construction, the depth of the original circuits corresponds to the depth of the PBC-compiled instances. As for the gate counts, the original circuits had respectively $18\,309\pm 219$ and $5\,875$ single- and 2-qubit gates, while the corresponding results for the PBC-compiled circuits were $2\,092\pm 129$ and $1\,475\pm 62$.

\subsection{Task 2: Virtual qubits and approximate strong simulation}

\begin{figure*}[t]
    \centering
    \begin{subfigure}[]{0.48\textwidth}
        \centering
        \includegraphics[scale=0.39]{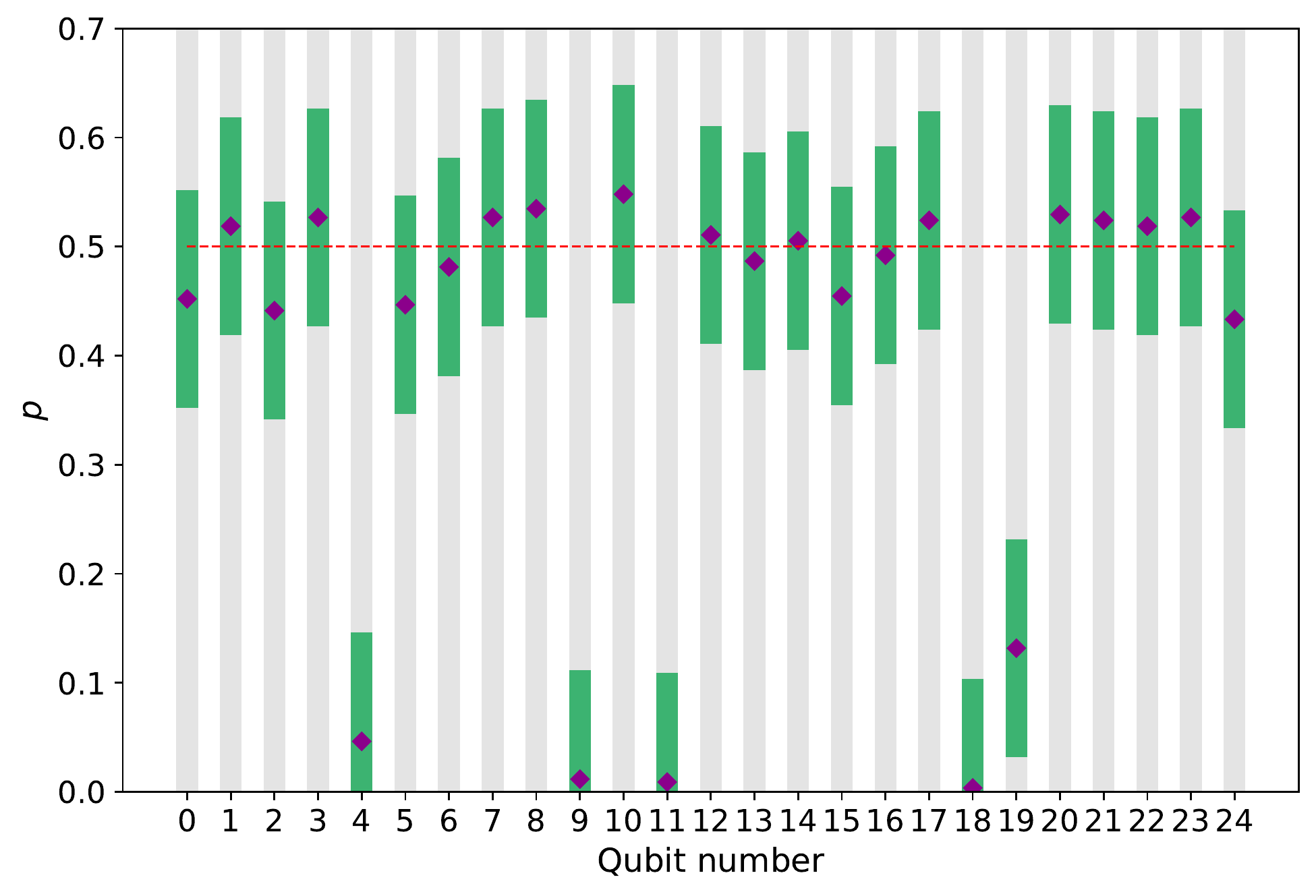}
        \caption{$\epsilon = 0.1$}
        \label{subfig: Hybrid-RQC0-0.1}
    \end{subfigure}
    \hfill
    \begin{subfigure}[]{0.48\textwidth}
        \centering
        \includegraphics[scale=0.39]{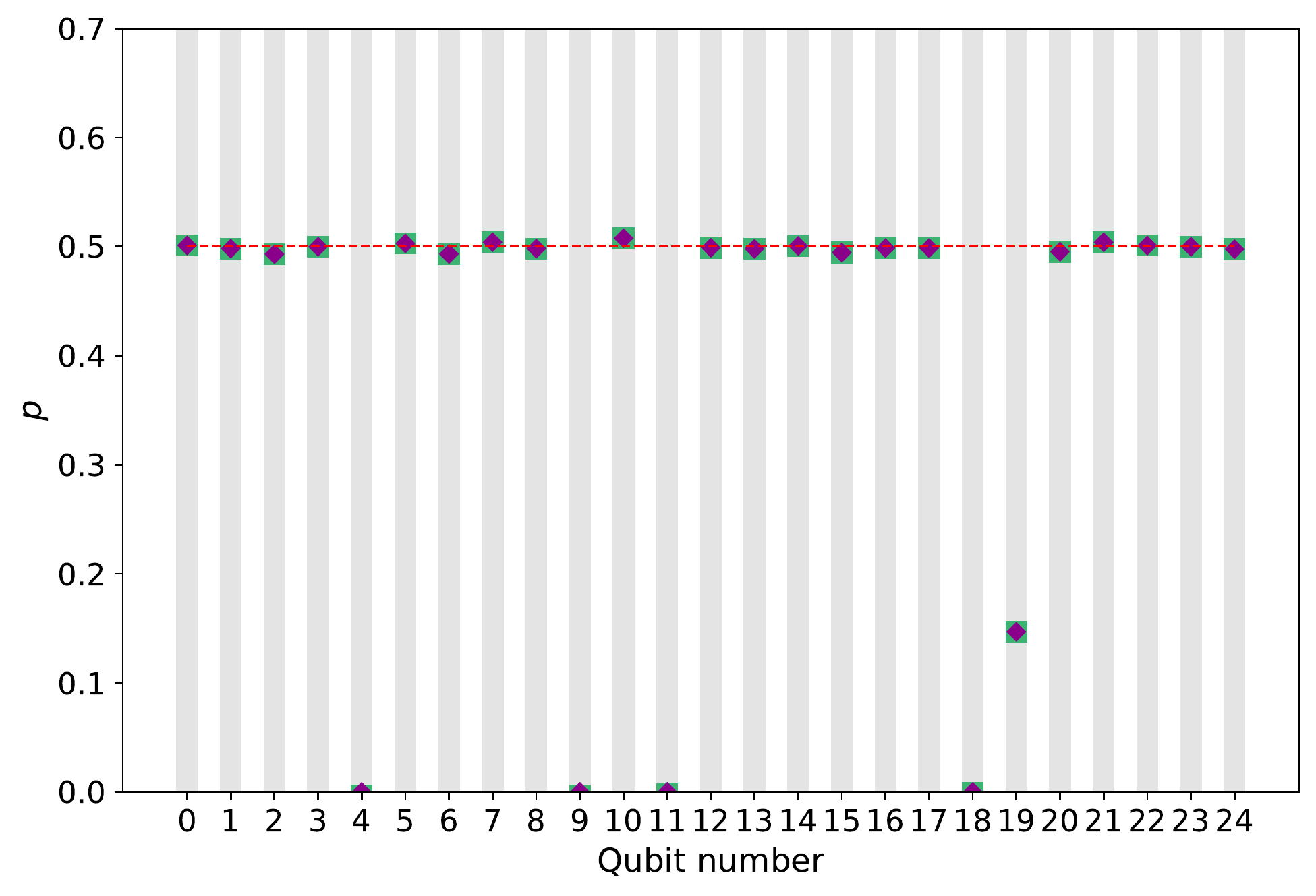}
        \caption{$\epsilon = 0.01$}
        \label{subfig: Hybrid-RQC0-0.01}
    \end{subfigure}
    \caption{Probability of getting the outcome 1 for each of the 25 qubits of a (randomly generated) random quantum circuit with 13 $T$ gates, for 2 different maximum error values: (a) $\epsilon = 0.1$, and (b) $\epsilon = 0.01$, and simulating a single virtual qubit. The total number of shots (per qubit) was $530$ and $52\,984$, respectively for $\epsilon = 0.1$ and $\epsilon = 0.01$ (cf. Table~\ref{tab: Hybrid_PBC-nr_samples}).
    The magenta diamonds represent the expectation value of the random variable $\eta$ and the green bars correspond to 99\% confidence intervals following Hoeffding's inequality.}
    \label{fig: Hybrid-RQC}
\end{figure*}

In this Section, we will present the results obtained for the hybrid PBC set-up, where we simulate different numbers of virtual qubits and determine the probability $p$ that a given qubit yields the outcome 1 with different precisions.

We start by considering one of the HSCs with $n=10$ belonging to the set of circuits used in Task 1. For this value of $n$, the hidden string is $\mathbf{s} = 1001110010$, so that each qubit has a deterministic output, that is, either $p=0$ or $p=1$. We expect that hybrid computation allows to estimate this probability within additive error $\epsilon$. For this particular case, we run the hybrid PBC with different numbers of virtual qubits, namely 1, 2, 3, and 4, and with a fixed, maximum additive error of $\epsilon = 0.1$ Additionally, we consider a 99\% confidence level so that the probability of failure in Equation~\eqref{eq: Improved sampling complexity hybrid} is $p_{fail}=0.01\,.$ In doing this, we require a total number of $\mathcal{N}=530,\,\, 1\,060,\,\, 2\,120,$ and $4\,239$ samples, respectively for 1, 2, 3, and 4 virtual qubits (cf. Table~\ref{tab: Hybrid_PBC-nr_samples}). The results are depicted in Figure~\ref{fig: Hybrid--HSC}. They illustrate how we can combine a classical machine and a small QPU to simulate a larger computation than allowed by the available quantum hardware. Specifically, we are simulating a PBC on 15 qubits via smaller PBCs involving, respectively, 14, 13, 12 and 11 qubits.

Evidently, if we decrease $\epsilon$ we will see a narrowing of the confidence intervals. We omit such an illustration for this particular circuit, as it will be done in the next case study.

As for Task 1, we would also like to illustrate this hybrid PBC setup for the case of input RQCs. The circuits considered in Section~\ref{subsubsec: RQCs} had $N_{cycles} = 40.$ For this reason, these circuits are very well mixed and their output probability distribution is (almost perfectly) uniform for nearly all instances. This means that $p\approx 0.5$ for (nearly) every qubit of each circuit.

To be able to run our hybrid PBC task and see deviations from this (near) uniform behavior we chose to generate a single random quantum circuit with $n=25,$ and $t=13$, but with only $N_{cycles} = 8.$ For that circuit, we estimated the output probability of each qubit using a Schr{\"o}dinger-type simulator and obtained the following results:

\begin{itemize}
    \item $p=0$ for qubits 4, 9, 11, and 18;
    \item $p\approx 0.146$ for qubit 19;
    \item $p \approx 0.5$ for all other qubits.
\end{itemize}
We expect to obtain the same results by carrying out hybrid PBC. We did so for two different maximum errors: $\epsilon = 0.1$ and $\epsilon = 0.01$ and only a single virtual qubit. Under these conditions, and considering again a 99\% confidence level, the total number of samples required to obtain the results was, respectively, $\mathcal{N}=530$ and $\mathcal{N}=52\,984$ (cf. Table~\ref{tab: Hybrid_PBC-nr_samples}). The results can be found in Figure~\ref{fig: Hybrid-RQC}.

As expected, the estimate of $p$ obtained using hybrid PBC is consistent with the values presented above. While these results were obtained when simulating a single virtual qubit, similar results would be obtained for larger virtual qubit counts.

\section{Concluding remarks} \label{sec: Concluding remarks}

We have shown how the Pauli-based model of quantum computation can be used in practice to compile a given unitary Clifford$+T$ circuit. Compared to prior work~\cite{Yoga19}, our first scheme (Section~\ref{subsubsec: Proposal 1}) reduces the number of quantum gates required to perform the computation, specifically, to $O(t^2)$ instead of $O(t^3/\log (t)),$ at the expense of requiring a single auxiliary qubit. Our second scheme (Section~\ref{subsubsec: 2nd proposed scheme -- t*logt depth}) reduces the computational depth from $O(t^2)$ to $O(t\log t)$ by using logarithmic-depth cascades of CNOTs akin to the constructions used for computing parity functions; it also avoids the auxiliary qubit needed in the first proposal. Finally, a space/time trade-off is possible; specifically, using $t$ auxiliary qubits allows us to reduce the depth of the PBC-compiled circuits to $O(t)$ (Section~\ref{subsubsec: 3rd proposed scheme -- linear depth}).

Besides these contributions, we also wrote Python code which effectively implements this compilation technique, demonstrating in practice the promised quantum resource minimization. To the best of our knowledge, this practical demonstration of the usefulness of PBC could not be found in the literature up to this point.

In the context of the considered hidden-shift circuits (HSCs), the resource minimization is clear from the observation of Figures~\ref{fig: HSCs--depths} and \ref{fig: Gate counts}. It is also evident for the larger instances with $n=t=42$ compiled by resorting to the dummy simulator as explained in Section~\ref{subsubsec: Dummies}.

In the case of the random quantum circuits (RQCs), the reduction in the number of quantum operations is still relevant (Figure~\ref{fig: RQCs -- Gate counts}), even if the depth of the computation increases with respect to that of the original computation for $t>7$. At this point, it is also worth noting that whenever the original computation is augmented only by Clifford operations, then the quantum resources required for the compiled computation (number of qubits, depth and gate counts) will remain essentially unchanged. As such, in the context of the RQCs considered, if we increased the number of cycles in the input circuits ($N_{cycles}>40$), we would still have the same results for the compiled computations, even though the original circuits would now be deeper and have even more single- and two-qubit unitaries. Put differently, our code becomes increasingly advantageous as the ratio $N_{cycles}/t$ increases.

Moreover, this indifference of the PBC-compilation technique to the addition of extra Clifford gates means that such operations can be regarded essentially as free, similarly to what happens in the one-way model of measurement-based quantum computation. Therefore, these models look very promising for the implementation of techniques or procedures wherein extra Clifford gates are applied in order to realize a given task. A straightforward example is the \textit{classical shadows} technique described in Ref.~\cite{HuangKP20}. Succinctly, this technique requires a number of copies of a certain (unknown) quantum state that scales as $O(\log M)$, in order to estimate $M$ properties of that state. The procedure relies on the application of Clifford unitaries, drawn randomly from a certain ensemble, to the unknown quantum state, before measuring it. These extra Clifford unitaries will only change the structure of the PBC to be computed, not the (overall) required resources like the number of qubits or the (average) depth and gate counts.

Finally, our work with hybrid PBC shows how we can extend the number of qubits of the quantum hardware by using a classical (super)computer to simulate $k$ virtual qubits, at the expense of having a cost exponential in $k$. We improve this exponential sampling complexity by a significant amount, with respect to prior work with a scaling of $O(3^k)$. Notably, we show that this quantity is upper bounded by $O(2^{0.7374k} \epsilon^{-2})$. Moreover, we also establish a lower bound for this sampling complexity of $\Omega (2^{0.5431k}\epsilon^{-2})$. By resorting to a small QPU, this set-up is asymptotically advantageous over the use of the best full classical stabilizer-rank simulators whenever the number qubits traded out is such that $k\lesssim t/2$. On the other hand, the use of quantum hardware inevitably implies the presence of coherent and incoherent noise, of which a full classical simulation is otherwise devoid.

There are several natural ways in which this work can be extended and improved, and thus be made even more useful. Firstly, we could adapt the code to allow adaptive Clifford$+T$ input quantum circuits. This would largely improve our results for the HSCs as we would be able to apply the same adaptive decomposition of the Toffoli gate as used in Refs.\cite{BG16} and \cite{BBCCGH19}, effectively reducing the $T$-count of each input circuit. In fact, and more interestingly, this advantage will extend to general circuits, as allowing adaptivity will enable the same computation to be encoded with a smaller number of $T$ gates, which is the factor that ultimately determines the depth and gate counts of the PBC-compiled circuits.

Secondly, also regarding the code itself, it could be fruitful to implement the second and third schemes for performing Pauli measurements (i.e., the schemes in Sections~\ref{subsubsec: 2nd proposed scheme -- t*logt depth} and \ref{subsubsec: 3rd proposed scheme -- linear depth}, respectively). This would allow more extensive comparative studies of the quantum resource savings and trade-offs achieved by PBC compilation. For instance, these schemes should bring significant improvements to the depth of the compiled circuits which, based on the numerical results in Section~\ref{subsec: Results - task 1}, should be particularly beneficial in the context of random quantum circuits.

We could also explore different pre-compilation techniques, which could be applied to the quantum circuit before running our code. In the context of strong simulation and hybrid PBC, a possibility might be to combine our code with the techniques developed in Ref.~\cite{KissWet21}, which sometimes allow for the cancellation of the non-Clifford phases in the ZX-diagrams.

Another potential line of research that might be worth pursuing relates to the presence of noise in real quantum hardware (briefly touched upon above). The hybrid setup provides not only a reduction of the number of qubits needed in the actual QPU but also guarantees that the virtual qubits are noise-free. In our current hybrid implementation, the code always removes the upper qubits; when thinking about noisy implementations, researching how the choice of different virtual qubits might improve the results in practice might prove significantly beneficial.

Finally, on a more engineering-related approach, changing this to a C++ or Julia implementation should reduce the classical processing time, lessening the demands on the coherence of the quantum hardware.

\section{Acknowledgements} \label{sec: Acknowledgements}
Circuit diagrams were created using Quantikz~\cite{Quantikz}. FCRP would like to thank Professor JMB Lopes dos Santos for fruitful discussions and Professor JM Viana P Lopes for providing access to the remote server wherein the code was run.
FCRP is supported by the Portuguese institution FCT – Funda\c{c}\~{a}o para a Ci\^{e}ncia e a Tecnologia via the PhD Research Scholarship 2020.07245.BD. EFG acknowledges funding of the Portuguese institution FCT – Funda\c{c}\~{a}o para a Ci\^{e}ncia e a Tecnologia via project CEECINST/00062/2018. This work was supported by the H2020-FETOPEN Grant PHOQUSING (GA no.: 899544).
The authors would also like to thank an anonymous reviewer for constructive and fruitful comments on the implementation of Pauli measurements and the assessment of the (in)dependence of Pauli operators.

\bibliographystyle{quantum}
\bibliography{Bib_paper}

\end{document}